\documentclass[
twocolumn,
superscriptaddress,
nofootinbib,
 amsmath,amssymb,
prstper,
floatfix,
]{revtex4-2}

\usepackage{graphicx}
\usepackage{dcolumn}
\usepackage{bm}
\usepackage{color,soul}
\usepackage[dvipsnames]{xcolor} 

\usepackage[utf8]{inputenc}

\begin{document}


\title{Percent Grade Scale Amplifies Racial/Ethnic Inequities in Introductory Physics}


\author{Cassandra A. Paul}
\affiliation{Department of Physics \& Astronomy - Science Education Program - San Jose State University}

\author{David J. Webb}
\affiliation{Department of Physics, University of California - Davis}


\date{\today}

\begin{abstract}
In previous work we analyzed databases for 95 classes to show that the percent grade scale was correlated with a much higher student fail rate than the 4.0 grade scale. This paper builds on this work and investigates equity gaps occurring under both scales.  By employing a ``Course Deficit Model'' we attribute the responsibility for closing the gaps to those who are responsible for the policies that guide the course.  When comparing course grades in classes graded using the percent scale with those in courses graded using the 4.0 scale, we find that students identifying as belonging to racial or ethnic minorities underrepresented in physics suffer a grade penalty under both grade scales but suffer an extra penalty under percent scale graded courses.  We then use the fraction of A grades each student earns on individual exam items as a proxy for the instructor's perception of each student's understanding of the course material to control for student understanding and find that the extra grade penalty students from groups underrepresented in physics students suffer under percent scale grading is independent of the student's understanding of physics.  When we control for more student level variables to determine the source of the grade scale dependent penalty, we find that it is primarily the low F grades (partial credit scores) on exam problems that are the source of these inequities.  We present an argument that switching from percent scale grading to a 4.0 grade scale (or similar grades scale) could reduce equity gaps by 20-25\% without making any other course changes or controlling for any incoming differences between students.

\end{abstract}


\maketitle

\title{RacismGradeScalePaper}
\author{cassandra.paul }
\date{December 2020}

\maketitle

\section{Introduction}

In our 2020 paper \cite{Webb2020a} we compared the grade distributions resulting from two different grade scales. We found that one grade scale, a version of percent grading, led to many more students failing than the other, a 4-point grade scale. In this paper we will follow up those general conclusions and show that the percent grade scale exerts a differentially negative effect on the grades from students who are members of racial/ethnic groups underrepresented in physics when compared to the grades of their peers. We also show that this differential effect does not seem to reflect differences in the students' understanding of physics. This result supports our current understanding that demographic grade gaps are largely the result of the policies and procedures of the course itself. Putting the onus for change on the course itself has been referred to by Cotner \& Ballen as a ``Course Deficit model'' \cite{cotner2017} of demographic gaps.  This name is to be contrasted with a ``Student Deficit model'' (see Ref. \cite{Valencia1997} by Valencia for a history of many of these kinds of models) of demographic gaps that uses student-level information in attempting to understand grade gaps at the group level. 

Our expectation that a course can and should deliver roughly equal results, on average, independent of demographic group membership has been called the ``Equity of Parity'' model of equity \cite{Rodriguez2012Impact}.  If a racial/ethnic group consistently receives a lower average grade than their peers then this model looks to changes in the course to rectify the inequity.  Racial/ethnic inequities that are caused by the course may be considered to be part of a larger system of structural racism\cite{Omi2015}.  Adopting a course deficit model for our explanation of a demographic gap is a natural result of expecting Equity of Parity.  This is because with an Equity of Parity model, any average differences in preparation or prior experiences between groups would not impact the groups' odds of being successful, thus any deviation from Equity of Parity is the fault of the course. It is also a particularly fruitful approach because attributing a demographic grade gap to the course may provide the instructor, who has control over the course, with the power to make changes toward achieving equity of parity. In contrast, the other prominent model of demographic grade gaps, the ``Student Deficit model'', attributes demographic inequities to student-level issues such as math and physics preparation. Lack of preparation presumably involves past inequities that are not easily dealt with by the instructor of a course. So the Course Deficit Model allows an instructor to see themselves as responsible for equity in their course, instead of solving any past inequities.

In comparing different demographic groups we will be discussing the differences in the average grades between the different demographic groups. These kinds of demographic gap studies (sometimes termed ``gap gazing'' \cite{Rodriguez2001}), have been subject to critiques recently \cite{Gutierrez2008}.  We share these concerns and note that a growing body of recent research can be used as evidence that these critiques are well taken. In the remainder of the introduction to this paper we will first outline some of the critiques of much of the literature on demographic grade gaps, connect that criticism to recent research, and outline how the data presented in this paper fits into the discussion.

\subsection{Gap Gazing and the Student Deficit Model}

In her critique of gap gazing \cite{Gutierrez2008}, Rochelle Guti\'errez points us to a set of interlocking issues as well as some technical problems with gap gazing.  First of three interlocking issues, the grade distributions of these different groups can be largely overlapping each other (e.g. studies \cite{Salehi2019}, \cite{Shafer2021}, and \cite{Webb2021a} show gaps less than half a standard deviation of the distributions), but the condensation of two broad distributions of grades down to two numbers, the averages of each group, brings the focus onto \textbf{differences} rather than similarities between two groups. One issue making this a problem is that each average grade is given the name of the entire group so some of the responsibility for the higher average grade would seem \cite{Quinn2020} to accrue to many students who actually had low grades, and some of the responsibility for the lower average grade would seem to accrue to many students who actually had high grades.  Second, and more importantly, over all of the history of these comparisons the group with the lower average has been thought of as having some kind of average deficit \cite{Valencia1997} (i.e. a Student Deficit Model is used). Taken together with the first point above, these deficits, if they existed, would then seem to accrue to all members of the group \cite{Quinn2020}.  Third, when researchers use the student deficit model in explaining grade gaps (for example by attributing the gaps to lack of preparation of certain groups \cite{Salehi2019}) then they tend not to provide instructors with any easy levers that can be used to change the situation. The course is completely under the instructors' control but they have nearly no control over who their students are. If we instead decide that the course itself is responsible for teaching all of the enrolled students and, hence, is responsible for the grade gap, then the instructors would have complete control over the levers necessary to affect the gap (although what changes to make to the course might still be a difficult question).

Often, in studies of demographic grade gaps researchers will rely on control variables that are well known to help in explaining grade differences \cite{Salehi2019,Hazari2007,Kost2009Characterizing} and in at least one case \cite{Salehi2019} going so far as to conclude that the control variables that explain the differences \textbf{within} groups also explain the the grade differences \textbf{between} groups.  Guti\'errez \cite{Gutierrez2008} also points out that this extension from within groups to between groups may not be valid. Indeed, in their 2021 paper Shafer et al. \cite{Shafer2021} have found that at least one common control variable, a student's SAT/ACT score, that is reliably positively correlated with within-group grades seem to be sometimes positively correlated with between-group grade gaps (Mexican-ancestry vs peers) but sometimes negatively correlated with between-group grade gaps (Asian-ancestry vs peers). In other words, Shafer et al. found that controlling for SAT/ACT scores (among other variables) decreases the equity gaps for Mexican-American students, but increases the equity gaps for Asian-American students. This may indicate that SAT/ACT scores are not good predictors of preparation or it may mean that student preparation itself is not useful in predicting between-group course grades. Within-groups vs between-group issues like this are also seen in Ref. \cite{Webb2021a} and make one wary of using a Student Deficit Model for understanding demographic gaps.

\subsection{The Course Deficit Model}

In defense of gap gazing, Lubienski argues that these types of studies are needed to illuminate ``which groups and curricular areas are most in need of intervention and additional study'' \cite{Lubienski2008}, and further claims that gap studies are necessary for advancing equitable policy changes. However, if we agree and see value in gap analyses, we must attend to the critiques made by Guti\'errez and others. Indeed, Guti\'errez argues not for the elimination of gap studies, but instead calls for more contextualized intervention studies \cite{Gutierrez2008}.  We see the application of the Course Deficit Model as a potential way of doing this. 

The authors of the current paper consider the course itself to be the problem leading to demographic gaps.  Here, ``course'' is shorthand for the material presented, the presentation itself (the order of ideas and practice and the pedagogical styles of lecture, discussion, and laboratory), the exams, and the grading.  Our conclusions in this regard are informed by the critiques of gap-gazing as well as by arguments made by writers such as Ta-Nehisi Coates \cite{Coates2014} and Ibram Kendi \cite{Kendi2019} who suggest that one should change the system which is responsible for perpetuating inequities and avoid blaming the victims of the inequities for suffering them. 

Recent education research supports our view and also suggests some changes in pedagogy \cite{Theobald2020,Webb2017,Webb2021a}, as well as other changes \cite{Webb2021} less connected to pedagogy, that might be made to repair an introductory physics course's deficiencies with respect to equity. The results in the current paper provide further support for the utility of a Course Deficit Model and add to the possible non-pedagogical changes in a path toward Equity of Parity. Specifically, we show how grading policies of a course can increase racial/ethnic grade gaps and how a common grade scale, the percent scale, amplifies this problem. 

We use the working hypothesis that the causes of the demographic grade gaps in physics courses are to be found in the organization and policies of the courses themselves and that no demographic group in our courses has any significant average incoming deficiency. Some evidence supporting this hypothesis is found in research by Webb \cite{Webb2017,Webb2021a} where students who were from underrepresented racial/ethnic groups (the American Physical Society recognizes female students as well as students from a set of racial/ethnic groups as underrepresented in physics) had higher final exam grades than their peers when they were in a class teaching concepts first rather than the much more common case, lower than their peers, that was found in the other three more traditionally organized classes taking the same final exam. Theobald et al. \cite{Theobald2020} also find that replacement of standard lecture courses with active-learning courses reduces grade gaps for underrepresented groups. A third example is found by Simmons et al. \cite{Simmons2020} who show that simply changing course grade weighting policies can lead to changes in the demographic grade gaps. Additionally Webb \cite{Webb2021} describes a change in the exam regimen within an active-learning introductory physics course which resulted in female identifying students receiving higher grades than male identifying students rather than the much more common case of lower grades.  These various studies, as well as other researchers \cite{Lubienski2008}, point out the usefulness of paying attention to demographic grade gaps, particularly when using a Course Deficit model and attending to changing a course to improve equity.  The present paper shows another change in grading policies that can lead to changes in demographic grade gaps.

\subsection{\label{sec:GradeScales}Previous work on CLASP grade scales}

The focus in our previous work has been on the actual grades students received because these actual grades have real effects on the students. A student's grade may determine whether they repeat the course, whether they change majors, and/or their self-efficacy within their major.  We will generally keep that focus in the present paper.

Our previous paper \cite{Webb2020a} was concerned with the two main grade scales used by instructors in the active-learning introductory physics series for biological science majors at UC Davis \cite{Potter2013Sixteen, Paul2017}.  We assembled the available grading databases for each class that was offered between 2003 and 2012, a period that included many classes where exams were graded on a 4-point scale (defined by us as CLASP4) and many classes graded on a 10-point percent-like scale (CLASP10). 

The grade scales are named this way because they were both used in the Collaborative Learning Through Active Sense-making in Physics (CLASP) curriculum. This course is an active-learning learning course, that consists of one 80 minute lecture per week (often including a weekly quiz) with the entire class of around 250 students, and two 2 hour and twenty minute discussion lab meetings per week of about 25 students that consist of students working in groups at white boards and with equipment in activity cycles that are entirely focused around small group and whole class discussions led by a teaching assistant (TA). While there are instructional style differences across the TAs the discussion labs are all highly interactive \cite{West2013Variation}.  For more details on the CLASP curriculum see reference \cite{Potter2014Sixteen}.

The course topics and course materials were almost identical over this set of years so that the main differences were the instructors and the grade scales. The grade scale was used for each problem on each exam and the exam score was calculated using a (sometimes weighted) average of the individual problem grades. Table \ref{tab:tab1} shows the letter grades associated with the numerical scores given under the two grade scales. The course grade was largely based on a weighted average of the student's exam scores so that the particular grade scale told each student how well they had done on each problem and, after averaging, on each exam and also, after averaging, what final course grade they could expect.

\begin{table}[htbp]
\caption{Comparing different grade scales.  The Letter Grade to Percent Scale and 4.0 Scale conversions are from the College Board website. CLASP10 is the specific version of the percent scale whose results are discussed in the paper and CLASP4 is the specific 4-point scale used. (The scales are named ``CLASP'' because they are used in the Collaborative Learning through Active Sense-making in Physics course studied in this paper.) }
\label{tab:tab1}
\begin{ruledtabular}
\begin{tabular}{c c c c c }
\textbf{Letter} & \multicolumn{2}{c}{\textbf{Common Scales}} & \multicolumn{2}{c}{\textbf{Specific Scales}} \\
\textbf{Grade} & Percent Scale & 4.0 Scale & CLASP10 & CLASP4 \\ 
 \hline
A+ & 97-100 & 4.0 & 9.67-10 & 4.17-4.5 \\
A & 93-96 & 4.0 & 9.33-9.67 & 3.83-4.17 \\
A$-$ & 90-92 & 3.7 & 9.0-9.33 & 3.5-3.83 \\
B+ & 87-89 & 3.3 & 8.67-9.0 & 3.17-3.5 \\
B & 83-86 & 3.0 & 8.33-8.67 & 2.83-3.17 \\
B$-$ & 80-82 & 2.7 & 8.0-8.33 & 2.5-2.83 \\
C+ & 77-79 & 2.3 & 7.67-8.0 & 2.17-2.5 \\
C & 73-76 & 2.0 & 7.33-7.67 & 1.83-2.17 \\
C$-$ & 70-72 & 1.7 & 7.0-7.33 & 1.5-1.83 \\
D+ & 67-69 & 1.3 & 6.67-7.0 & 1.17-1.5 \\
D & 65-66 & 1.0 & 6.33-6.67 & 0.83-1.17 \\
D$-$ &  &  & 6.0-6.33 & 0.5-0.83 \\
F & 0-65 & 0.0 & 0-6.0 & 0-0.5 \\
\end{tabular}
\end{ruledtabular}
\end{table}

As we have shown in our previous work, a student's actual grade may have considerable dependence on the grading scale that an instructor chooses in organizing their course. For instance, instructors using a percent scale gave 5 times as many course grades less than C- than those using a 4-point scale \cite{Webb2020a}.  In grading individual questions on exams, the percent scale instructors also gave many more nonzero F-grades. Because these two results were true for each instructor who used both grade scales, these conclusions seem to be grade scale dependent but independent of instructor. Finally, we showed that the large increase in course grades less than C- under the percent grade scales were mainly the result of the heavy effective weight of low F-grades on exam questions/problems when averaging under percent grade scales. So zeros, received due to leaving answers blank and missing exams, are most important but other nonzero F's (for example scoring 3 points out of a possible 10) are also important.

\section{\label{sec:Methods}Methods}

\subsection{\label{sec:FracAs}Fraction of A-grades is a \\measure of understanding}

To compare the effects of grade scale on different demographic groups we pick a control variable that i) is highly correlated with the course grade, ii) is consistent across grade scales (both theoretically and empirically), and iii) we consider to be a proxy for each student's understanding of the course material.  

Our database has each grade given to each student by that student's instructor on each individual exam problem answer that that student gave. Those grades are certainly related to the student's demonstrated understanding of the appropriate physics material. We propose choosing a subset of these grades, the A grades, and using the fraction of A grades received as a metric that serves as a proxy for that student's physics understanding. This fraction certainly satisfies point i) above because the correlation between Course Grade and Fraction of A-grades is r = 0.83 where anything above r = 0.75 is generally considered a strong correlation.  We will argue that it satisfies ii) and iii) and has some other useful qualities as well.  Importantly, we are not arguing that the fraction of A-grades is necessarily a better metric of understanding than course grades on either scale, but we will make the case that it is a useful proxy for understanding, one that is not impacted by the low grades on either scale which are the source of the difference between the two scales.

The campus grading rules tell us that an A-grade denotes ``excellence'' so a particular student’s fraction of A-grades should be proportional to their instructor’s opinion of the fraction of the course material in which that student has demonstrated excellent understanding.  This is close to a prima facie case that this fraction is one possible measurement of demonstrated physics understanding.  While we agree with critiques that grades do not always indicate understanding \cite{Wieman2005}, we argue that the fraction of A-grades a student earns is related to the instructors' perception of how much a student has mastered and so should satisfy point iii) above. We will call this ``understanding of the material'' as shorthand for ``the instructor's perception of the student's understanding of the material.''  Regarding point ii) above, we have already noted that A grades have a meaning, excellent, that is theoretically independent of grade scale and we will examine the actual grades across grade scales later in the paper to verify that this measure is appropriate in practice as well as in theory.

In addition, the fact that these grade come from the classes themselves gives them some characteristics that we should note. First, the fraction of A-grades will likely exhibit whatever racial/ethnic bias that is found in the course grades. For this reason, one might have hoped that using the fraction of A-grades as a control variable will also control for all of the racial/ethnic bias of the course. Unfortunately we will show that there is additional bias that is accounted for by grading practices.  Second, in our previous paper \cite{Webb2020a} we showed that there is considerable class-to-class variation in course grades that is not easily attributable to class-to-class variation in the academic abilities of the students. We also showed that this large grade variation, which we could call ``grade noise'', has both a between-instructor part and a within-instructor part. We may well be able to filter out much of this grade noise by using a control variable, fraction of A grades, whose distribution is distinct to each class and so will somehow also include the same grade noise. Finally, at this point we should note that all of our general conclusions will still hold if we had chosen a narrower definition of ``understanding'' by confining it to A+ answers or a broader definition to include both A and B-graded answers.

\subsection{\label{sec:CLASP}Database used in this study}

In using a student's fraction of A's as a measure of their physics understanding we decided to limit our database to the classes for which we had all of the exam grades (all quiz/midterm grades and all final exam grades) so that the fraction of A's was an accurate measure of understanding of the entire course.  For this same reason, we also only included students who i) received a course grade and ii) had grades on at least 50\% of the graded exam items.  We found 73 class databases that included all of the exam-item grades given to the students.  We added university-supplied relevant demographic data (i.e. gender, racial/ethnic group identity, first generation status, and citizenship) to complete the database we use for this study.  In our consideration of racial/ethnic issues we decided to follow the APS definitions \cite{APS} and remove from consideration the 2\% of the students who are neither US citizens nor permanent residents.  Our final database included 11,047 students in 49 CLASP4 classes and 5,574 students in 24 CLASP10 classes.  For the CLASP4 classes 12\% of these students are members of racial/ethnic groups (African heritage, Latin American heritage, Mexican heritage, Native American heritage, and Pacific Islander heritage) identified by APS as underrepresented in physics and for the CLASP10 classes that number is 13\%.

\subsection{\label{sec:StatMethods}Variables and Statistical methods}

We will compare student course grades ($CourseGrade$) under the two specific grade scales, CLASP4 and CLASP10 (see Table \ref{tab:tab1}) using the fraction of A-grades a student receives ($FracAs$) to control for student understanding of the material.  As in our previous paper \cite{Webb2020a} we use the UC Davis numerical values for course grade given  by A = 4.0,
A- = 3.7, B+ = 3.3, B = 3.0, etc. except that we use A+ = 4.3 rather than the UC Davis A+ = 4.0.  We compare the effects of the two grade scales on all students and also the differential effects on students from racial/ethnic groups historically underrepresented in physics. Fitting $CourseGrade$ vs $FracAs$ for the individual classes shows us that this function usually has a small negative curvature so we include both the linear term ($FracAs$) and the quadratic term ($FracAs2$) when we control for physics understanding.\footnote{ We note that using just the linear term $FracAs$ does not change our numbers much and does not change our conclusions at all.} We use the categorical variable $PercentScale$ (= 0 for CLASP4 and = 1 for CLASP10) to measure the average shift downward of CLASP10 grades when compared to CLASP4 grades for students with equal physics understanding. Our previous work \cite{Paul2018} showed large racial/ethnic differences in some behaviors (leaving answers blank and/or not taking all quizzes) associated with course grade so we want to examine the effects of these behaviors on the grades.  To do this we use the categorical variable $URM$ (= 0 when student does not identify as belonging to a racial/ethnic group underrepresented in physics as defined by APS \cite{APS}, and = 1 when the student does belong to a so-defined underrepresented racial/ethnic group) in our models. In addition, we expect the effects of these behaviors on course grade to depend on grade scale so we will include an interaction term between grade scale and $URM$ status to allow us to measure the average effect of the grade scale on these students.

$FracAs$ and $URM$ are variables that vary from student to student within each class while the categorical variable $PercentScale$ is the same for each student in a class and only varies from class to class. In addition, we expect other class-to-class differences (e.g. an instructor may average all quizzes together or drop each student’s lowest quiz or drop the two lowest quizzes) will cause students’ grades to be correlated by class rather than independent at the student level. Ordinary least-square (OLS) regression assumes uncorrelated errors so, for the models we will fit, we expect OLS to compute incorrect standard errors, possibly leading to incorrect statistical inferences.  Because of this issue with OLS we will use Hierarchical Linear Modeling (HLM).  HLM will account for these kinds of class-dependent correlations, when predicting $CourseGrade$, by modeling each class as a group when finding the best overall fits to our models.  Specifically, in using $FracAs$ as a control variable the linear and quadratic parts are treated as fixed variables and we account for class-to-class differences by allowing the constant term to be a random class-dependent variable.  For more detailed information on HLM see reference \cite{VanDusen2019}.  We also check the HLM results against OLS regression to make sure we understand any differences.  A small issue with HLM is that there is no absolute measure, like $R^2$ for OLS, of how well the model fits the data.  Snijders \& Bosker \cite{Snijders1994} offer an alternative calculation for $R^2$, analogous to what is used in OLS.  We will use the Snijders/Bosker $R^2$ ($SBR^2$) determined for each level and note that, in each model we fit, the student-level value of $SBR^2$ is slightly smaller than the $R^2$ we would have gotten if we had used OLS for the same model.

\section{\label{sec:Results}Results}

\subsection{\label{sec:A-gradeData}A-grades under different grade scales}

Because actual instructors may award A-grades differently under the different grade scales we will check the average number of A-grades given in our current data set for evidence of differences between the two grade scales.  Overall, instructors using CLASP4 grading gave A's on exam problems/questions 40.7\% $\pm$ 0.5\% of the time while under CLASP10 grading the percentage of A-grades was 45.0\% $\pm$ 0.6\%.  For this difference Cohen's d is 0.17 which, under the standard usage for Cohen's d, is a small effect.  This comparison can be contrasted with the much larger difference in F-grades given under the two grade scales.  The CLASP4 instructors gave non-zero F-grades 3.9\% $\pm$ 0.1\% of the time while under CLASP10 the percentage of non zero F's was considerably larger at 17.9\% $\pm$ 0.5\%.  For this difference Cohen's $d$ is 1.7 which, under the standard usage for Cohen's $d$, is certainly a large sized effect.  This comparison suggests that the percentage of A-grades given is relatively stable when switching grade scales.

A more general model allowing us to directly compare the two grade scales is likely to require us to take into account the fact that the students are grouped into classes.  One reason this grouping seems important is that exam difficulty is almost certainly instructor-dependent and the grading itself may be instructor-dependent. We can check these possibilities as well as test the grade-scale dependence of the fraction of A-grades using HLM with the model shown in equation \ref{eqn:HLMModel0}.   \begin{multline}
FracAs = b_0 \\+ b_{PercentScale}PercentScale + b_{URM}URM \\+ b_{URM*PercentScale}(URM*PercentScale)
\label{eqn:HLMModel0}
\end{multline}

\begin{table}[htbp]
\caption{The coefficients from an HLM fit to equation \ref{eqn:HLMModel0} are shown along with their standard errors, z-statistics, and P-values.  Included are $N=16,621$ students in 73 classes.}
\label{tab:tab1a}
\begin{ruledtabular}
\begin{tabular}{c c c c c}
\textbf{Coeff.} & \textbf{Value} &\textbf{Error} & \textbf{z-statistic}
& \textbf{P-value}\\ 
\hline
$b_{PercentScale}$ & -0.016 & 0.019 & 0.83 & 0.409 \\
$b_{URM}$ & -0.051 & 0.005 & -10.69 & $<10^{-3}$ \\
$b_{URM*PercentScale}$ & -0.003 & 0.008 & -0.36 & 0.718 \\
$b_{0}$ & 0.414 & 0.011 & 37.91 & $<10^{-3}$ \\
\end{tabular}
\end{ruledtabular}
\end{table}

The coefficients from the HLM fit to equation \ref{eqn:HLMModel0} are given in Table \ref{tab:tab1a}.  From $b_0$ we find the overall fraction of A's given to non-$URM$ student groups is 41.4\% and, as expected, we find that the fraction of A's shows the same racial/ethnic inequity that one often finds for course grades, $b_{URM} = -0.051 \pm 0.005$ with P $< 10^{-3}$.  In other words, regardless of which grade scale is used in this study, there are some inherent inequities in how A's are awarded, and equity of parity is not met for this metric for any number of reasons. Nevertheless, because it represents the amount of material the instructor perceives a student has mastered, it is useful to control for this variable to understand the differential impacts of the grade scale itself. For the present study comparing the two grade scales the more relevant results are that the effect of the grade scale on the fraction of A-grades received is statistically insignificant for both the $URM$ student groups ($b_{URM*PercentScale} = 0.003 \pm 0.008$ so P = 0.72) and their peers ($b_{PercentScale} = 0.016 \pm 0.019$ so P = 0.41).\footnote{We can also note that no other set of letter grades, neither B's, C's, D's, nor F's, is given equally under the two grade scales.}  This lends more evidence to the claim that fraction of A's is independent of grade scale. Along with these two parameters we also find that, within this model, the between-class variance in fraction of A-grades is a little over 20\% of the within-class (student-to-student) variance.  We can reduce the within-class variance by controlling for incoming student GPA but this does not reduce the between-class variance.  The large between-class variance suggests that there is a large instructor/exam/grading effect.  Whatever the cause, this considerable class-to-class variation tells us that the hierarchical structure of the data has to be taken into account in our statistical inferences about whether fraction of A-grades was dependent on grade scale.  We conclude that for our purposes the average fraction of A-grades given is statistically independent of the grade scale used by the instructor. 

\begin{figure}
\includegraphics[trim=3.6cm 3.0cm 5.2cm 3.1cm, clip=true,width=\linewidth]{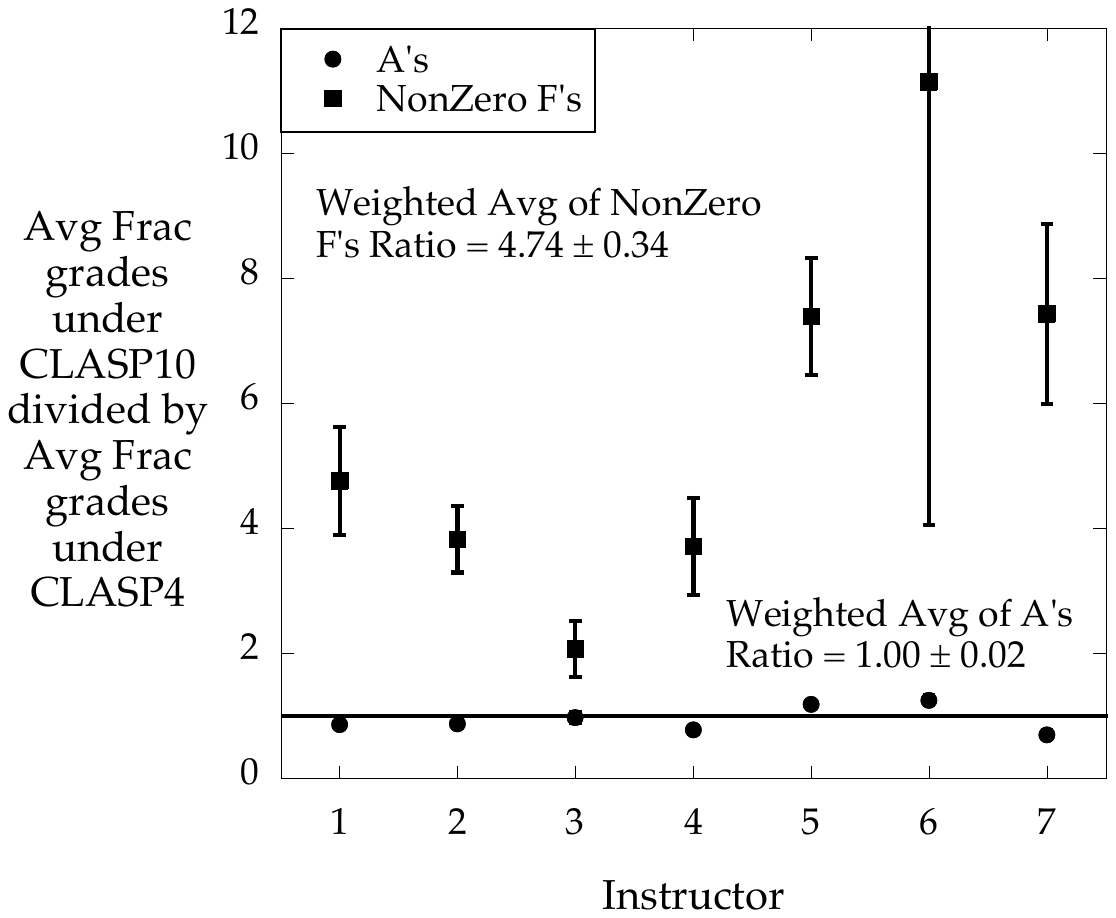}
\caption{Each of seven instructors taught under both grade scales.  The figure shows both the ratio of their class-averaged fraction of A-grades under CLASP10 and that fraction under CLASP4 and the similar ratio for non-zero F-grades.  A horizontal line is drawn at a ratio of 1 where grades are awarded equally under the two scales.  We see that these instructors treated A-grades roughly the same under the two grade scales even though they treated non-zero F-grades very differently.  The error bars are estimated standard errors.}
\label{Fig2}
\end{figure}

A final possible issue with these grade scale comparisons is that any instructor effect is convolved with the grade scale effect. We can do a similar grade scale comparison that does not have an instructor dependence by examining data from the seven instructors who used both grade scales at various times. For each of these seven instructors we find the average fraction of A's given under each grade scale and then plot the ratio = (average fraction of A-grades under CLASP10) / (average fraction under CLASP4). If an instructor treats grades the same under the two grade scales then this fraction will equal one.  As seen from Figure \ref{Fig2}, these instructors gave essentially the \textbf{same} fraction of A-grades under CLASP4 as they did under CLASP10 (the average ratio is 1.00 $\pm$ 0.02). For comparison, they gave almost 5 times as many non zero F-grades under CLASP10 as under CLASP4. Given all of these various ways of looking at this issue it seems fair to conclude that the number of A-grades given has little connection to the grade scale chosen by the instructor. Hence our confidence in using this fraction as a control variable, internal to each class, that allows us to compare these two grade scales. 

\subsection{\label{sec:URMResult}Differential effects for\\ historically excluded racial/ethnic groups}

\subsubsection{\label{sec:URMOverall}Grade gaps for historically excluded groups}

First, we compare the treatment of $URM$ student groups under the two grade scales. We use HLM to fit equation \ref{eqn:HLMModel1} which includes a grade scale term, a $URM$ demographic term, and the interaction between these two. The resulting coefficients are shown in Table \ref{tab:tab2}. The grade scale coefficient shows the overall downward skewing of a percent scale like CLASP10. The $URM$ coefficient shows that CLASP4 graded students from underrepresented groups receive grades of almost a quarter of a grade point less than their peers. This equity gap is a relatively common literature result \cite{Salehi2019, Shafer2021, Webb2021a} but, as we have noted in the introduction, applying an Equity of Parity model leads us to expect that all demographic groups on average will achieve statistically indistinguishable outcomes.  Thus we take this grade gap as indicating a deficiency in the course. Finally, the interaction term shows that the CLASP10 grade scale increases the grade gap by over 30\% more than the CLASP4 base level. We now try to understand these effects better by controlling for the instructor's perception of each students' physics understanding using a student's fraction of A-grades.

\begin{multline}
CourseGrade = b_0 \\+ b_{PercentScale}PercentScale + b_{URM}URM \\+ b_{URM*PercentScale}(URM*PercentScale)
\label{eqn:HLMModel1}
\end{multline}

\begin{table}[htbp]
\caption{The coefficients from an HLM fit to equation \ref{eqn:HLMModel1} are shown along with their standard errors, z-statistics, and P-values.  Included are $N=16,621$ students in 73 classes.  We see an overall $URM$ grade penalty of 0.225 with an additional penalty under percent-scale grading of 0.075.  At the student level this model has $SBR^2 = 0.11$ with $SBR^2 = 0.03$ at the class level.}
\label{tab:tab2}
\begin{ruledtabular}
\begin{tabular}{c c c c c}
\textbf{Coeff.} & \textbf{Value} &\textbf{Error} & \textbf{z-statistic}
& \textbf{P-value}\\ 
\hline
$b_{PercentScale}$ & -0.168 & 0.058 & -2.89 & 0.004 \\
$b_{URM}$ & -0.225 & 0.020 & -11.00 & $<10^{-3}$ \\
$b_{URM*PercentScale}$ & -0.075 & 0.035 & -2.17 & 0.030 \\
$b_{0}$ & 2.925 & 0.033 & 87.52 & $<10^{-3}$ \\
\end{tabular}
\end{ruledtabular}
\end{table}

\subsubsection{\label{sec:URMControl}Grade gaps even after controlling for understanding}

We again compare the grade scales but now we use the fraction of each student's answers that were judged as excellent to control for that student's understanding of physics in doing our comparison.  Again, HLM is used in fitting our data to equation \ref{eqn:HLMModel2} which includes the fraction of A's both in a linear term and a quadratic term for the reasons discussed earlier.  The resulting coefficients are shown in Table \ref{tab:tab3}.  The $SBR^2$ associated with this model shows that it is an excellent fit at the student level (since the model includes student grades). On the other hand, the class level is not well explained by our model.  This isn't surprising to us because, as we showed earlier \cite{Webb2020a}, there is considerable class-to-class variation.  In this paper we are just interested in the overall effects of the grade scales on the students and not on modelling the class-to-class variation.

\begin{multline}
CourseGrade = b_0 \\+ b_{FracAs} FracAs + b_{FracAs2} FracAs^{2} \\+ b_{PercentScale}PercentScale + b_{URM}URM \\+ b_{URM*PercentScale}(URM*PercentScale)
\label{eqn:HLMModel2}
\end{multline}

\begin{table}[htbp]
\caption{The coefficients from an HLM fit to equation \ref{eqn:HLMModel2} are shown along with their standard errors, z-statistics, and P-values.  One sees that the $URM$ grade penalty due to percent-scale grading (compared to 4-point scale grading) is about the same after controlling for physics understanding as it was without that control.  At the student level this model has $SBR^2 = 0.70$ with $SBR^2 = 0.16$ at the class level.}
\label{tab:tab3}
\begin{ruledtabular}
\begin{tabular}{c c c c c}
\textbf{Coeff.} & \textbf{Value} &\textbf{Error} & \textbf{z-statistic}
& \textbf{P-value}\\ 
\hline
$b_{FracAs}$ & 4.923 & 0.062 & 79.40 & $<10^{-3}$ \\
$b_{FracAs2}$ & -1.341 & 0.067 & -19.89 & $<10^{-3}$ \\
$b_{PercentScale}$ & -0.224 & 0.057 & -3.97 & $<10^{-3}$ \\
$b_{URM}$ & -0.031 & 0.010 & -3.10 & 0.002 \\
$b_{URM*PercentScale}$ & -0.067 & 0.017 & -3.96 & $<10^{-3}$ \\
$b_{0}$ & 1.158 & 0.035 & 33.16 & $<10^{-3}$ \\
\end{tabular}
\end{ruledtabular}
\end{table}

As above the coefficient $b_{PercentScale}$ shows that the percent-like CLASP10 grade scale skews students grades downward on average.  The surprise (to us) is that even after controlling for physics understanding a student from a racial/ethnic group underrepresented in physics still receives, on average, a lower grade under CLASP4 grading ($b_{URM} = -0.031$) \textbf{and} an additional lower grade under CLASP10 (percent scale) grading ($b_{URM*Percentscale}= -0.067$).  This extra grade penalty under percent scale grading is roughly the same size as it was before controlling for understanding.  Under CLASP10 this amounts to a total $URM$ grade penalty of about $0.10 \pm 0.02$ grade points that are not explained in terms of the students' understanding of the subject.  We should point out that one gets the same results just using ordinary multivariable regression except that, as expected, for ordinary regression the error in the grade scale coefficient is (inappropriately) much smaller.

Finally, for many of the databases we have not only all of the grades but the computation of the exam grade with instructor-determined weights for the quizzes and the final exam (including possibly dropping one or more low quizzes).  This exam grade is by far the most important part of the course grade but, as discussed in reference \cite{Webb2020a}, there are small grade adjustments determined by discussion/lab participation, lecture participation, homework, etc.  We have used HLM for the model in equation \ref{eqn:HLMModel2} with ``$ExamGrade$'' substituted for ``$CourseGrade$'', to our data and find essentially the same grade penalties for $URM$ groups.  This tells us that the grade penalties are due to exams and not to other parts of the course.

\subsubsection{\label{sec:AllEthn}Grade penalties/advantages for several ethnicities}

\begin{figure}
\includegraphics[trim=3.2cm 2.8cm 5.5cm 3.3cm, clip=true,width=\linewidth]{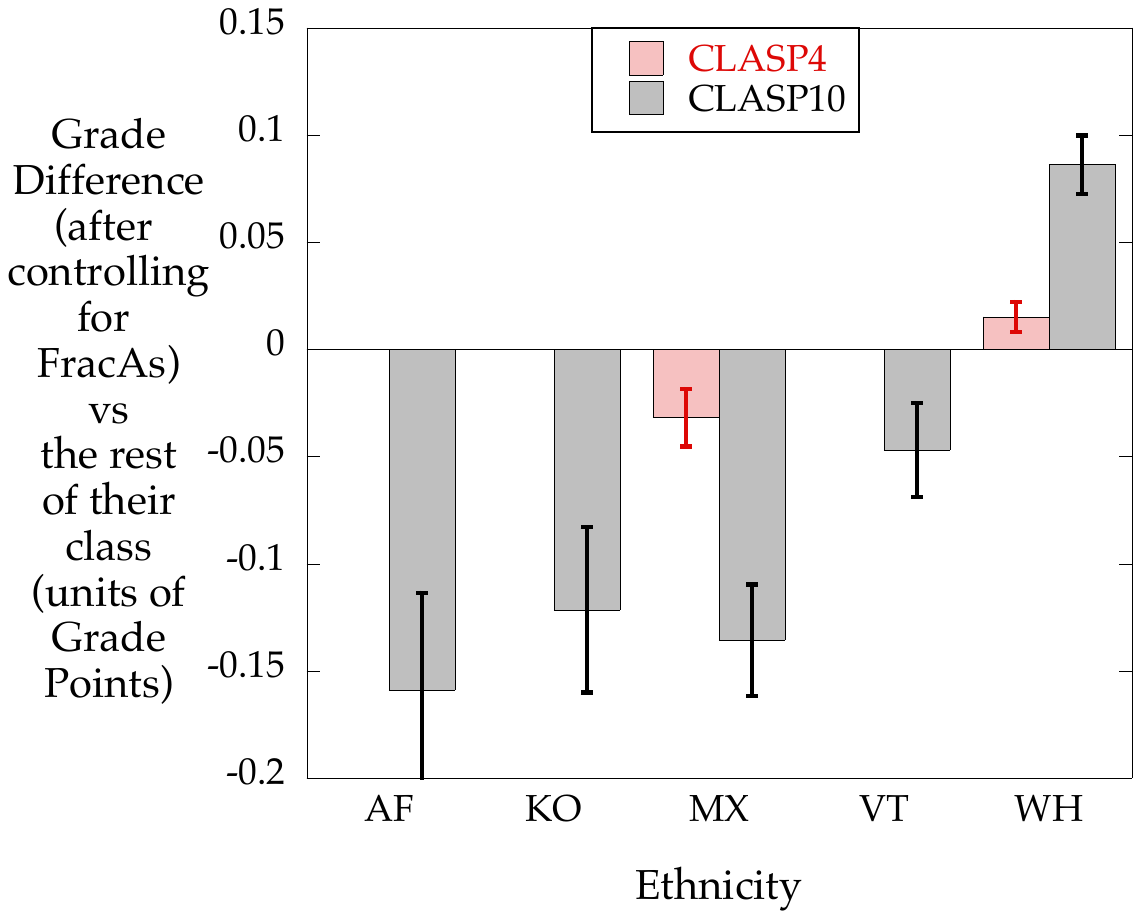}
\caption{The grade penalties from Table \ref{tab:tab4} that are statistically significant at the $P<0.05$ level.  AF, KO, MX, VT, and WH refer to the group of students with African, Korean, Mexican, Vietnamese, and white (causasian), respectively, heritage.  A negative value shows that the relevant group had lower grades than their peers in the class after controlling for physics understanding using their fraction of A-grades. The error bars are estimated standard errors.}
\label{Fig2a}
\end{figure}

Reference \cite{Shafer2021} showed that aggregating several different ethnicities together (as we do with $URM$) in one's analyses can lead to loss of relevant information regarding the impact that the course might have on large groups of students.  We examine this as a possible issue in this grade scale study by not only disaggregating the group $URM$, but by comparing each individual racial/ethnic group (defined for us by our administration) against all of their peers in the same way that we did for $URM$.  Specifically, for the twelve identified racial/ethnic groups in our classes we use twelve individual HLM models shown generically as
\begin{multline}
CourseGrade = b_0 \\+ b_{FracAs} FracAs + b_{FracAs2} FracAs^{2} \\+ b_{PercentScale}PercentScale + b_{Eth}Eth \\+ b_{Eth*PercentScale}(Eth*PercentScale)
\label{eqn:HLMModel2a}
\end{multline}
to measure each group against all of their peers in the class while controlling for physics understanding.  For each racial/ethnic group $b_{Eth}$ represents the grade penalty under 4-point grades scales, CLASP4, and $(b_{Eth} + b_{Eth*PercentScale})$ represents the grade penalty under the percent grade scales, CLASP10.  The results of these twelve models are shown in Table \ref{tab:tab4}.  Noting that only a few of these results are statistically significant at the level of $P<0.05$, we plot those significant results in Figure \ref{Fig2a}.  Note that this is not the TOTAL grade penalty but just the part of the penalty that is related to race/ethnicity after controlling for student understanding.

First, we note that about 52\% of the students from $URM$ groups in our sample were of Mexican heritage so, given our $URM$ results, it isn't surprising that these students received a grade penalty which was amplified by percent scale grading. The group of students in this data set who are of Latin American (but not Mexican) heritage did not have a statistically significant grade penalties but students of African heritage had approximately the same penalties as those of Mexican heritage. So, as seen in the conclusions of reference \cite{Shafer2021}, breaking the $URM$ group down into smaller identifiable groups does change our results slightly.  Nevertheless, our conclusions about percent scale grading amplifying the average grade penalties given to some racial/ethnic groups who are underrepresented in physics are not changed.  Some subgroups of $URM$ student groups did receive grade penalties that were likely unrelated to their understanding of physics.

Second, any conclusions we would draw for students whose ancestry is asian would depend on the country their ancestors are from.  The largest number of these students have Chinese ancestry and it is quite clear that they receive essentially zero penalty compared to all of their peers in the class.  On the other hand, students with Korean heritage receive grade penalties of about the same size as those with Mexican or African heritage after controlling for physics understanding.

Finally, we find that students with white (caucasian) heritage constitute a group who are privileged under the percent scale, receiving an extra grade advantage that has little relation to their understanding of physics.  Of course, since the grade penalty applied to each group is measured against the rest of their class, if some demographic groups receive grade penalties compared to their peers then others must be receiving grade advantages when compared to their peers from the groups with grade penalties.

\begin{table}[htbp]
\caption{The grade penalties (along with their standard errors and P-values), measured in grade points, are shown for each grade scale for each of the twelve ethnicities identified in these classes.  A separate HLM model, Eq. \ref{eqn:HLMModel2a}, is used for each ethnicity.  Thus, the grade bias comparing each ethnic group to the rest of their class, after controlling for fraction of A-grades, is shown.  These are not the total grade penalties suffered by these groups but, instead, the grade penalty after controlling for physics understanding using the student's fraction of A-grades on exam answers.  * indicates $0.01<P<0.05$, ** indicates $0.001<P<0.01$, and *** indicates $P < 0.001$.}
\label{tab:tab4}
\begin{ruledtabular}
\begin{tabular}{c c c c c}
\textbf{Heritage} & \textbf{Grade Scale} & \textbf{Penalty} &\textbf{Error} & \textbf{P-value}\\ 
\hline
African & CLASP4 & -0.019 & 0.022 & 0.391 \\
African & CLASP10 & -0.159*** & 0.046 & $<10^{-3}$ \\
\hline
Chinese & CLASP4 & -0.0004 & 0.0081 & 0.96 \\
Chinese & CLASP10 & 0.008 & 0.016 & 0.63 \\
\hline
East Indian & CLASP4 & 0.0001 & 0.0142 & 0.99 \\
East Indian & CLASP10 & -0.027 & 0.028 & 0.34 \\
\hline
Filipino & CLASP4 & 0.018 & 0.013 & 0.18 \\
Filipino & CLASP10 & -0.035 & 0.028 & 0.21 \\
\hline
Japanese & CLASP4 & -0.004 & 0.024 & 0.88 \\
Japanese & CLASP10 & 0.035 & 0.047 & 0.45 \\
\hline
Korean & CLASP4 & -0.033 & 0.020 & 0.1 \\
Korean & CLASP10 & -0.122** & 0.039 & 0.002 \\
\hline
Lat. Amer. & CLASP4 & -0.015 & 0.023 & 0.51 \\
Lat. Amer. & CLASP10 & -0.012 & 0.043 & 0.78 \\
\hline
Mexican & CLASP4 & -0.032* & 0.013 & 0.018 \\
Mexican & CLASP10 & -0.136*** & 0.026 & $<10^{-3}$ \\
\hline
Native Amer. & CLASP4 & -0.027 & 0.046 & 0.56 \\
Native Amer. & CLASP10 & 0.035 & 0.086 & 0.68 \\
\hline
Other Asian & CLASP4 & -0.002 & 0.018 & 0.91 \\
Other Asian & CLASP10 & -0.020 & 0.034 & 0.56 \\
\hline
Pacific Isl. & CLASP4 & -0.040 & 0.029 & 0.17 \\
Pacific Isl. & CLASP10 & 0.123 & 0.076 & 0.11 \\
\hline
Vietnamese & CLASP4 & 0.007 & 0.011 & 0.52 \\
Vietnamese & CLASP10 & -0.047* & 0.022 & 0.033 \\
\hline
White & CLASP4 & 0.0152* & 0.0069 & 0.027 \\
White & CLASP10 & 0.086*** & 0.014 & $<10^{-3}$ \\
\end{tabular}
\end{ruledtabular}
\end{table}

\subsection{\label{sec:URMExplan}Why is this grade gap present?}

Since the extra grade penalty given to $URM$ student groups under percent grading (CLASP10) is not obviously due to physics understanding it seems worthwhile to track down its origin. First we'll show that it is not likely that instructors using CLASP10 are grading $URM$ students differently from those using CLASP4 and then explore the likely reasons for the basic $URM$ grade penalty and then for the additional CLASP10 grade penalty.

\subsubsection{\label{sec:OvertRacism?}Grade scale dependence of grade gap is not an instructor effect}

In our previous paper \cite{Webb2020a} we noted that a major difference between CLASP4 and CLASP10 is that the grade distribution resulting from CLASP10 was over 30\% wider, that this broadening was largely due to the heavy effective weight given to the low F-grades when averaging under percent grading, and that this broadening effect was independent of instructor.  We can remove the broadening effect by normalizing the $CourseGrade$ distributions (to produce $ZCourseGrade$) so that every class has an average grade of zero with a standard deviation of 1.  We similarly normalize, for each class, the distribution of $FracAs$ (giving $ZFracAs$) so that we can model how one distribution is mapped into the other and whether that mapping depends on the grade scale or on a student's identification as a member of an underrepresented group.  We have already argued that the fraction of A-grades is a measure of understanding of the physics and is similar for the two grade scales so we expect that the mapping between these two distributions is roughly independent of grade scale and hope that any shift of $URM$ student groups in this mapping will be independent of the grade scale.

First, we compare grade scales.  The model we fit is:
\begin{multline}
ZCourseGrade = b_0 + b_{ZFracAs} ZFracAs \\+ b_{ZFracAs2} ZFracAs^{2} + b_{PercentScale}PercentScale
\label{eqn:HLMModel3}
\end{multline}

For this model we find that the effect due to the CLASP10 grade scale (i.e. the coefficient $b_{PercentScale}$) is -0.0005 $\pm$ 0.0074 (P = 0.950).  The very small effect size and large error estimate tells us that there is no distinguishable grade scale effect in the mapping from the normalized fraction of A's to the normalized course grade.

To finish the discussion of these normalized distributions we'll examine whether the mapping between normalized distributions preserves the discriminatory grade penalties that we found in the actual distributions.  Although we don't expect a grade scale effect, we will still include it and its interaction with $URM$.  In other words, we will fit the following:
\begin{multline}
ZCourseGrade = b_0 + b_{ZFracAs} ZFracAs \\+ b_{ZFracAs2} ZFracAs^{2} + b_{PercentScale}PercentScale \\+ b_{URM}URM + b_{PercentScale*URM}(PercentScale*URM)
\label{eqn:HLMModel4}
\end{multline}

The coefficients that we find from this fitting procedure are shown in Table \ref{tab:tab5}.  The $URM$ grade penalty is still statistically significant but, importantly, we see that there are no significant percent scale effects left (P-value is 0.69 for $b_{PercentScale}$ and P-value is 0.26 for $b_{URM*PercentScale}$) which shows us that the instructors from the two grade scales are treating their students roughly equally.  So, when we remove the grade-scale-dependence from the course grade distributions by normalizing them we also remove the grade-scale-dependence of the $URM$ grade gap.  This suggests that the grade scale difference in the $URM$ grade gap has the same origin as the grade scale difference \cite{Webb2020a} in the course grade distributions.

\begin{table}[htbp]
\caption{The coefficients from from an HLM fit  to equation \ref{eqn:HLMModel4} are shown along with their standard errors, z-statistics, and P-values.  The basic URM grade penalty is present even though we have normalized both the classes distributions of grade and of fraction of exam-item A-grades given.  As expected the grade scale effects are both statistically insignificant after these normalizations.}
\label{tab:tab5}
\begin{ruledtabular}
\begin{tabular}{c c c c c}
\textbf{Coeff.} & \textbf{Value} &\textbf{Error} & \textbf{z-statistic}
& \textbf{P-value}\\ 
\hline
$b_{ZFracAs}$ & 0.9080 & 0.0036 & 249.7 & $<10^{-3}$ \\
$b_{ZFracAs2}$ & -0.0617 & 0.0028 & -22.09 & $<10^{-3}$ \\
$b_{PercentScale}$ & -0.0031 & 0.0079 & -0.39 & 0.693 \\
$b_{URM}$ & -0.077 & 0.013 &  -5.88 & $<10^{-3}$ \\
$b_{URM*PercentScale}$ & 0.025 & 0.022 & 1.14 & 0.255 \\
$b_{0}$ & 0.0699 & 0.0053 & 13.12 & $<10^{-3}$ \\
\end{tabular}
\end{ruledtabular}
\end{table}

\subsubsection{\label{sec:MissingData}Missing data gives rise to part of the gap}

Our previous work \cite{Webb2020a} has shown that the main differences between the CLASP4 and CLASP10 grade scales result from the extra effective weight of low F-grades under CLASP10 (perhaps coupled with the many more non-zero F-grades given to students under CLASP10).  The extra weight given to these low grades is largest for grades of zero when a student leaves an answer blank or misses an exam. We have shown that leaving an answer blank or missing an exam seems to be a group-dependent behavior with $URM$ groups leaving more answers blank and missing more exams than non-$URM$ groups on average \cite{Paul2018}.  In searching for the origin of the $URM$ grade penalty we will first attempt to control for these issues of giving a grade of zero when an answer is left blank or an exam is missed (i.e. when grading data are missing). To that end we define two new variables.  $Frac0s$ is the fraction of a student's answers which received 0 because the student did not try to answer.  $FracMissQs$ is the fraction of quizzes missed by a student.  We use the same HLM fitting procedure with the following model:

\begin{multline}
CourseGrade = b_0 \\+ b_{FracAs} FracAs + b_{FracAs2} FracAs^{2} \\ + b_{PercentScale}PercentScale + b_{URM}URM \\ + b_{URM*PercentScale}(URM*PercentScale) \\ + 
b_{Frac0s}Frac0s +
b_{FracMissQs}FracMissQs \\
\label{eqn:HLMModel5}
\end{multline}

The coefficients that we find from this fitting procedure are shown in Table \ref{tab:tab6}.

\begin{table}[htbp]
\caption{The coefficients from an HLM fit to equation \ref{eqn:HLMModel5} are shown along with their standard errors, z-statistics, and P-values.  Using the fraction of blank answers and the fraction of missed quizzes as control variables we find that the basic (CLASP4) $URM$ grade penalty, $b_{URM}$, has been reduced to nearly zero but the extra percent scale grade penalty, $b_{URM*PercentScale}$, has not significantly changed.  At the student level this model has $SBR^2 = 0.76$ with $SBR^2 = 0.35$ at the class level.}
\label{tab:tab6}
\begin{ruledtabular}
\begin{tabular}{c c c c c}
\textbf{Coeff.} & \textbf{Value} &\textbf{Error} & \textbf{z-statistic}
& \textbf{P-value}\\ 
\hline
$b_{FracAs}$ & 4.132 & 0.059 & 70.06 & $<10^{-3}$ \\
$b_{FracAs2}$ & -0.824 & 0.063 & -13.16 & $<10^{-3}$ \\
$b_{PercentScale}$ & -0.234 & 0.051 & 4.55 & $<10^{-3}$ \\
$b_{URM}$ & -0.0019 & 0.0091 &  -0.21 & 0.832 \\
$b_{URM*PercentScale}$ & -0.058 & 0.016 & -3.71 & $<10^{-3}$ \\
$b_{Frac0s}$ & -2.793 & 0.074 & -37.76 & $<10^{-3}$ \\
$b_{FracMissQs}$ & -1.109 & 0.029 & -38.46 & $<10^{-3}$ \\
$b_{0}$ & 1.504 & 0.032 & 47.42 & $<10^{-3}$ \\
\end{tabular}
\end{ruledtabular}
\end{table}

The $b_{URM}$ coefficient from this model shows us that controlling for leaving blanks and missing quizzes leaves us with a CLASP4 grade penalty consistent with zero (i.e. reduced by a factor of 16 and no longer significantly different from zero grade penalty). Nevertheless, our suspicion that these issues might also explain the extra percent-scale (CLASP10) grade penalty does not appear to be born out.  The interaction term coefficient, $b_{URM*PercentScale}$, of -0.058 is still over 85\% of the value it had before our attempt to correct for missing data.  In other words, \textbf{controlling for the instructor's perception of student understanding (FracAs), and the fraction of missing work (Frac0s, FracMissQs) explains the grade penalty for $URM$ groups in CLASP4 courses, but does not explain the grade penalty for $URM$ groups in courses graded using the CLASP10 grade scale.}  If missing data do not explain the percent-scale $URM$ grade penalty then we need to look elsewhere.  We know that percent grading gives extra effective weight to low F-grades when averaging and that percent scale graders gave many more non-zero F-grades, so these may be the grades leading to the percent-scale's extra grade penalty.

\subsubsection{\label{sec:PercentSkewing}Percent-scale skews grades downward}

\begin{figure}
\includegraphics[trim=1.9cm 1.7cm 1.9cm 1.9cm, clip=true,width=\linewidth]{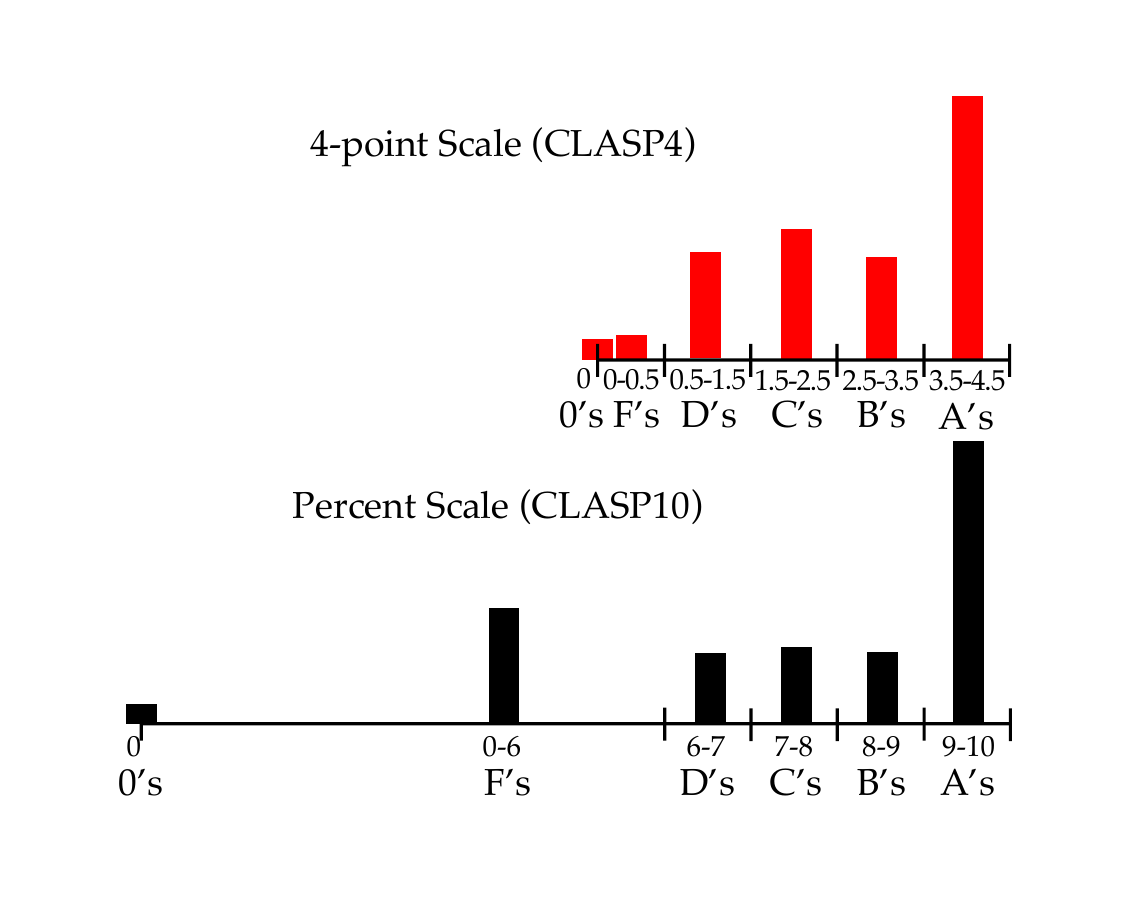}
\caption{The fractions of exam-item grades given under the two grade scales, CLASP4 and CLASP10, are shown.  The non-zero F-grades under CLASP10 are placed at the average value of these grades.  Note that there are many more non-zero F-grades given under CLASP10 \textbf{and} that the average F-grade is lower than \textbf{any} F-grade under CLASP4 so they carry a large weight when averaging and tend to skew the grade distribution downward.}
\label{Fig3}
\end{figure}

Figure \ref{Fig3} shows the exam-item grade distributions under the two grade scales we are comparing in this paper.  In our previous paper \cite{Webb2020a} we showed that percent-scale grading skewed student's grades downward and that the heavy effective weight given to low F-grades led to this skewing.  The figure shows that these many low F-grades under CLASP10 do not correspond to any grades available to instructors using CLASP4 and leads us to suggest that controlling for these F-grades may account for the extra $URM$ grade penalty under percent grading.  

To test this idea we include another variable, $FracFs$ = a student's fraction of non-zero F-grades, in the previous model.  So now we use HLM to fit the following:
\begin{multline}
CourseGrade = b_0 \\+ b_{FracAs} FracAs + b_{FracAs2} FracAs^{2} \\ + b_{PercentScale}PercentScale + b_{URM}URM \\ + b_{URM*PercentScale}(URM*PercentScale) \\ + 
b_{Frac0s}Frac0s +
b_{FracMissQs}FracMissQs \\ + 
b_{FracFs}FracFs
\label{eqn:HLMModel6}
\end{multline}

The coefficients that we find from this fitting procedure are shown in Table \ref{tab:tab7}.  Finally we have included enough variables so that there is no statistically significant $URM$ grade penalty beyond those resulting from the missing data (blanks and missed quizzes) and the percent-scale grading whose F-grades are given a large effective weight.  So we conclude that these three things produce the bulk of the grade penalty given disproportionately to students from underrepresented groups. 

\begin{table}[htbp]
\caption{The coefficients from an HLM fit to equation \ref{eqn:HLMModel6} are shown along with their standard errors, z-statistics, and P-values.  Including the fraction of non-zero F-grades as a further control variable we find that the additional percent-scale $URM$ grade penalty, $b_{URM*PercentScale}$, is reduced enough that it is no longer significant (in the statistical sense).  At the student level this model has $SBR^2 = 0.74$ with $SBR^2 = -0.08$ at the class level.}
\label{tab:tab7}
\begin{ruledtabular}
\begin{tabular}{c c c c c}
\textbf{Coeff.} & \textbf{Value} &\textbf{Error} & \textbf{z-statistic}
& \textbf{P-value}\\ 
\hline
$b_{FracAs}$ & 3.0951 & 0.056 & 55.67 & $<10^{-3}$ \\
$b_{FracAs2}$ & -0.357 & 0.058 & -6.29 & $<10^{-3}$ \\
$b_{PercentScale}$ & 0.211 & 0.067 & 3.16 & $<10^{-3}$ \\
$b_{URM}$ & -0.0116 & 0.0081 &  -1.43 & 0.153 \\
$b_{URM*PercentScale}$ & -0.013 & 0.014 & -0.91 & 0.361 \\
$b_{Frac0s}$ & -3.029 & 0.067 & -45.46 & $<10^{-3}$ \\
$b_{FracMissQs}$ & -1.054 & 0.026 & -40.66 & $<10^{-3}$ \\
$b_{FracFs}$ & -2.840 & 0.046 & -61.90 & $<10^{-3}$ \\
$b_{0}$ & 1.953 & 0.039 & 49.32 & $<10^{-3}$ \\
\end{tabular}
\end{ruledtabular}
\end{table}

We also note that the coefficient for the grade scale has changed sign.  This might seem unusual but our examination of the databases and the grade calculations show us that instructors using percent-scale grading often include in their calculations some things that will increase their students' grades.  Examples of the things some percent-scale instructors (but no 4-point scale instructors) have included are i) dropping each student's \textbf{two} lowest quizzes, ii) rescaling the final exam to increase all students' final exam grades, and iii) adding the same small fraction of grade points to each student's numerical grade before computing a letter grade.  Each of these things will raise the class-averaged grade but none of these were present in any 4-point graded databases.  We expect that these things that various percent-scale instructors have done lead to the positive value of $b_{Percent Scale}$ after we have controlled for all of the negative-skewing effects of the low F-grades.  We also note that controlling for these low F-grades in our model reduces $SBR^2$ at both the class level and the student level.  This issue is well known \cite{Snijders1994} for a model variable that behaves differently at the student level (lowers course grade) than at the class level (may induce instructor to raise all grades) so we are cautious about over-interpreting the coefficients at this point.

\subsection{\label{sec:URMHowmany}Effects of Grade Penalties on students}

There are two main impacts on students caused by percent-grading's wider grade distribution. First, as we discussed in our previous paper, there are many more students who receive grade less than C$-$ \cite{Webb2020a}.  Relatedly, there are many more students who must take the course again and so delay their academic careers. We will address both of these issues in terms of equity for students from underrepresented groups.  

The fraction of grades in our dataset that are below C$-$ is shown for both grade scales in Table \ref{tab:tab8} both for students from underrepresented racial/ethnic groups and for their peers.  The difference between these two percentages represents the percentage of the group that we estimate would have received a passing grade ($>$D$+$) under 4-point grading but would have received a failing grade ($<$C$-$) under percent grading.  Using the numbers in the table, we find that students from underrepresented groups are 83\% $\pm$ 24\% more likely to be pushed from a passing grade under CLASP4 to a failing grade under CLASP10 than are their peers.  

\begin{table}[htbp]
\caption{Percentage of grades less than C$-$ given to each group of students in our dataset.  We give results for each grade scale and also the number that would seem to be shifted from passing (above a D$+$) under CLASP4 to failing (below a C$-$) under CLASP10.  Standard errors are shown in parentheses.}
\label{tab:tab8}
\begin{ruledtabular}
\begin{tabular}{c c c c}
\textbf{Group} & \textbf{CLASP4} &\textbf{CLASP10} & \textbf{\% shifted from}\\
\textbf{} & \textbf{\% $<$C$-$} &\textbf{\% $<$C$-$} & \textbf{CLASP4 passing}\\ 
\textbf{} & \textbf{} &\textbf{} & \textbf{to CLASP10 failing}\\
\hline
$URM$ & 2.60 (0.43) & 14.2 (1.3) & 11.6 (1.4) \\
Non$URM$ & 0.89 (0.10) & 7.22 (0.37) & 6.33 (0.38) \\
\end{tabular}
\end{ruledtabular}
\end{table}

Next we turn to the related issue of a student deciding to repeat a course.  When students are given a grade less than C$-$ they are allowed to repeat the class so some of them will repeat it.  The option to repeat is a complicated issue, which has both positive and negative aspects. On the one hand, repeating a course and then being successful the second time is positive assuming that the student gained understanding and/or skills from that experience. On the other hand, taking a course a second time sets students back in their academic career. In other words, repeating a course is good if the repetition prepares you for later success, but if it does not do that, it is a waste of the students' time and the university's resources. In this section we share data regarding how the grade scales impacts the students' chances of repeating the course for two reasons: 1) to see how the grade penalty is impacting the course trajectories of $URM$ and $NonURM$  students, and 2) to frame the results in the following section where we discuss student grades in later coursework.

The fraction of instances of a student repeating a course from our dataset is shown for both grade scales in Table \ref{tab:tab9} both for students from underrepresented racial/ethnic groups and for their peers.    The difference between these two percentages represents the percentage of the group that we estimate would not have repeated the course under 4-point grading even though they did repeat under percent grading, and those numbers are included in Table \ref{tab:tab9} as well. Using the numbers in the table, we find that students from underrepresented groups are 59\% $\pm$ 29\% more likely to be pushed from not repeating a class under CLASP4 to repeating the class under CLASP10 than are their peers.

\begin{table}[htbp]
\caption{Percentage of student repeats after taking a course from our dataset.  We give results for each grade scale and also the number that would seem to be shifted from not repeating a course under CLASP4 to repeating the course under CLASP10.  Standard errors are shown in parentheses.}
\label{tab:tab9}
\begin{ruledtabular}
\begin{tabular}{c c c c}
\textbf{Group} & \textbf{CLASP4} &\textbf{CLASP10} & \textbf{\% shifted from}\\
\textbf{} & \textbf{\% repeats} &\textbf{\% repeats} & \textbf{CLASP4 not repeating}\\ 
\textbf{} & \textbf{} &\textbf{} & \textbf{to CLASP10 repeating}\\
\hline
$URM$ & 0.89 (0.26) & 6.63 (0.92) & 5.7 (1.0) \\
Non$URM$ & 0.35 (0.06) & 3.97 (0.28) & 3.62 (0.28) \\
\end{tabular}
\end{ruledtabular}
\end{table}

\subsection{\label{sec:LaterCourses}Student Grades in Later Coursework}

Upon finishing the CLASP series the students in our database completed an average of 75 (quarter) units of coursework to finish their undergraduate careers.  This is approximately 1.6 years of coursework.  One might worry that teachers who use a CLASP4 grade scale will pass students who are not prepared for some of this future coursework and who would have been forced to repeat a physics course (and so better prepare themselves) under a CLASP10 grading regime.  Examination of the student's GPAs in their later work gives us no evidence of this problem.  In Table \ref{tab:tab10} we show these after-physics GPAs for all students and also for students from underrepresented racial/ethnic groups.  For each grade scale we have separated the results according to the GPA they had upon entering their CLASP course.  We would have the same conclusion if we didn't break the groups up by their prior GPAs and also if we broke them up into even finer prior-GPA gradations so it seems clear that the later work of the CLASP4 students who passed their course but would have had to repeat it under CLASP10 is not noticeably different than those who did repeat physics.  In other words, the cost to the students when an instructor uses a percent scale do not seem to be balanced by any benefits to the students.

\begin{table}[htbp]
\caption{GPAs of student work after they have completed the CLASP physics series.  These are broken down by grade scale in their class as well as by the GPA they had upon entering CLASP.  Standard errors are shown in parentheses.}
\label{tab:tab10}
\begin{ruledtabular}
\begin{tabular}{c c c c c}
\textbf{Group} & \textbf{GPA} &\textbf{GPA} &\textbf{GPA} & \textbf{P}\\
\textbf{} & \textbf{before} &\textbf{after} &\textbf{after} & \textbf{(t-test}\\ 
\textbf{} & \textbf{CLASP} & \textbf{CLASP4} & \textbf{CLASP10} & \textbf{after)}\\
\hline
Whole & 2 to 3 & 2.79 (0.01) & 2.78 (0.01) & 0.65 \\
class & 3 to 4 & 3.41 (0.01) & 3.39 (0.01) & 0.36 \\
\hline
$URM$ & 2 to 3 & 2.76 (0.02) & 2.71 (0.04) & 0.19 \\
only & 3 to 4 & 3.31 (0.03) & 3.30 (0.03) & 0.86 \\
\end{tabular}
\end{ruledtabular}
\end{table}

One might wonder if there are benefits to the percent scale that only accrue to those students who actually repeated the course and received a higher grade than their first grade.  We have this list of students but we have no way of definitively choosing the comparison group, the students from 4-point scale classes who would have failed under percent grading.  However, our previous work \cite{Webb2020a} on rescaling grades in some of these classes suggests that over 90\% of $<$C$-$ percent scale students who would have succeeded under 4-point grading would have ended up with grades of C$-$ or C under 4-point grading.  So we will compare i) students from percent scale classes who then repeated the class with a higher grade to ii) students from 4-point scale classes who received C$-$ or C grades.  This comparison is shown in Table \ref{tab:tab11} where we see that there are no statistical differences between these two groups.  These conclusions don't change if we control for the students' incoming GPAs.  So, again, there is no obvious net benefit to a student repeating the course who wouldn't have had to repeat it under 4-point grading.

\begin{table}[htbp]
\caption{GPAs of student work after they have completed the CLASP physics series.  We compare students from percent scale graded classes who repeated the course and received higher grades with students from 4-point scale classes who received grades C$-$ or C and so could not repeat the course.  Standard errors are shown in parentheses.}
\label{tab:tab11}
\begin{ruledtabular}
\begin{tabular}{c c c c}
\textbf{Group} &\textbf{GPA after} &\textbf{GPA after} & \textbf{P}\\
\textbf{} &\textbf{C$-$ or C} &\textbf{successful repeat} & \textbf{(compare}\\ 
\textbf{} & \textbf{CLASP4} & \textbf{CLASP10} & \textbf{t-test)}\\
\hline
Whole class & 2.63 (0.01) & 2.58 (0.04) & 0.18 \\
\hline
$URM$ only & 2.64 (0.03) & 2.58 (0.10) & 0.50 \\
\end{tabular}
\end{ruledtabular}
\end{table}

\section{\label{sec:Discussion}Discussion}

\subsection{{\label{sec:Summary}Summary}}

Our previous work shows that the percent scale causes more students to fail when compared to the 4.0 scale \cite{Webb2020a}. We calculate this penalty here to be about 0.2 grade point average (GPA) points (see $b_{PercentScale}$ Tables \ref{tab:tab2} \& \ref{tab:tab3}).  Our main new finding is that students from racial/ethnic groups underrepresented in physics received a larger grade penalty than their peers under percent scale grading than under 4-point scale grading (see $b_{URM*PercentScale}$ from Table \ref{tab:tab2}).  We find the percent scale's differentially larger grade penalty  i) remains even after controlling for understanding using the students' fraction of A's as a control (Table \ref{tab:tab3})  ii) leads to many more of this group of students repeating the course (Table \ref{tab:tab9}) despite iii) there being no obvious benefit to them from this course repetition (Table \ref{tab:tab10}).  Also, when separated out by student identified ethnicity, in every case where there is a statistically significant penalty, the grade penalty is negative except in the case of white identifying students (see Table \ref{tab:tab4}).  This penalty (or bonus in the case of white students) is in addition to the penalty all students suffer under percent scale grading.  A Course Deficit Model \cite{cotner2017} identifies this extra grade penalty as an inequity that results from the course's grading practices. We find that the extra penalty results largely from the much larger fraction of F-grades that percent scale graders gave to their students' exam answers than did 4-point scale graders (see Table \ref{tab:tab6}).  The fact that those F-grades carry more effective weight under percent grading (when averaging to produce a course grade) than F-grades given under 4-point grading is the main reason for the large inequity. This leads us to suggest that grading reform - in particular a move from traditional percent scale grading to a 4.0 scale grading - is a partial solution to Equity of Parity grade gaps. The usefulness of the Course Deficit Model in helping to fix inequities, tempts us to call it the ``Course Improvement Model'' or the ``Course Empowerment Model''.

\subsection{{\label{sec:Implications}Implications}}

The use of a Course Deficit Model allows the instructor to concentrate on fixing inequitable outcomes resulting from their course. In assigning the source of inequities to the course itself, we situate it as the cause of inequities, but more importantly, as a possible solution to decreasing those same inequities. In this particular course sequence, we find the use of the percent scale to be an inequitable policy. We recommend switching back to the more equitable 4.0 scale, or trying other alternative grading methods shown to improve equitable outcomes.

In discussions of this work with faculty peers, the authors have found that some instructors are hesitant to abandon the percent scale because they see the 4.0 scale as `easier grading' and are concerned that they are doing students a disservice by passing students who are unprepared for later coursework. We hope these concerns are partially assuaged by the analysis conducted in Section \ref{sec:LaterCourses} which shows that no measurable difference between the average GPAs of students graded with the 4.0 scale and percent scale.  

We would also argue against using an unequal system to prepare students.  Our analysis (in Section \ref{sec:PercentSkewing}) showing that the low F grades are a primary source of the grade penalty indicates that under the percent scale students are weighted more heavily by their failures than their successes and that this penalty unequally impacts students who identify as belonging to racial and ethnic identities underrepresented in physics even after controlling  for the students' understanding. If more instructors embraced a Course Deficit Model, we would not need to prepare students for future inequities. Said simply, we do not support perpetuating inequities for the sake of preparing students to exist in an inequitable system.

In our analysis we find that $URM$ groups suffer grade penalties from both scales. We suspect that the reason for these inequitable penalties is that different demographic groups may, on average, differ in what they write on exams when they are unsure of their answer. In prior work we have already noted one of these differences: distinctly different numbers of blank answers given by different demographic groups \cite{Paul2018}.  Traditional thinking might classify these behavior differences as a deficiency of the student, and attempt to rectify the situation by providing supports to help the students overcome these test-taking behavioral deficiencies.  Indeed, this potential solution is also shared by faculty peers when we discuss this work.  However, employing a Course Deficit Model would, instead, attempt to remove the \textbf{impact} of the behavioral differences, noting that the instructional system privileges one type of behavior over another, thus inequitably impacting students of different demographics.  

We argue that instructors are likely aware that their percent scale grading is too harsh on their students.  We find that the instructors using the percent scale commonly do things to increase their students' grades (e.g. re-scaling their final exam to increase all students' grades, or adding the same number of points to each students' score) and that these are instructor behaviors are not found with those instructors using the 4.0 scale.  

We have previously shown \cite{Webb2020a} that the lower-graded 50\% of the answers (F through C+) have a very different grade distribution under percent grading than under 4-point grading. Percent graders gave students' exam answers F-grades roughly ten times as often as did 4-point graders (with a concomitant decrease in C and D grades under percent scales). These large differences that seem strongly influenced by the grade scale itself led us to conclude that this 50\% of the exam solutions are judged rather subjectively. Research shows that instructors award a large range of grades when awarding partial credit \cite{Henderson2004Grading, Marshman2017}.  It is in exactly these judgments of C, D, and F grades that the extra grade-scale-dependent penalties for $URM$ groups is found. In other words, perhaps unsurprisingly, it appears that racial/ethnic bias results from the most subjective grading.  It is beyond the scope of this paper to investigate the mechanism behind this particular anomaly, however, we suspect that the grade-scale-dependent penalties are due to structural racism which favors certain test-taking strategies over others. We guess this as an extension of the evidence we have previously shown of a demographic group dependence of other test-taking strategies, such as the choice of leaving an answer blank \cite{Paul2018}.  However, we acknowledge that other mechanisms, such as implicit bias, are not easily ruled out.

Employing a Course Deficit Model may require a paradigm shift on the part of the instructor. For example, when using an Equity of Parity Model, Rodriguez et al. argue, ``one must acknowledge that the instruction benefits the `less prepared' students more than the `well prepared' students'' \cite{Rodriguez2012Impact}.  For those instructors who have been using an Equity of Fairness Model \cite{Rodriguez2012Impact} (a model that ignores responsibility for past inequities by focusing on all demographic groups achieving the same gain) this might be a challenging change of perspective.  However, those using an Equity of Fairness Model also should recognize that to truly achieve this model of equity \textbf{requires} perpetuating inequities.  In shrinking our focus to what is under our control, we as instructors can begin to change the systematic inequities in higher education by holding our courses to a standard that does not further perpetuate inequities. 

With the results that we share here we argue that switching away from percent scale grading will remove a measurable amount of course grade inequity in courses that currently suffer from equity gaps.  We also argue that using a Course Deficit Model together with Equity of Parity can be one useful strategy among many in working towards more equitable education.

\subsection{{\label{sec: Future}Limitations and Avenues for Future Research}}

Perhaps the most important limitation of this study is that when using a Course Deficit Model we are only addressing equity as it manifests in the course, and even then we are  limited to addressing the inequities we actually measure. We are not solving structural racism in our classrooms, nor are we addressing it at the societal level. This is not to say that the model entirely ignores past inequities, rather it assumes that whatever inequities might exist should be inconsequential to success in the course. Furthermore, in our particular application of this model, we only address equity of student ``achievement'' as measured by grades.  Guti\'errez \cite{Gutierrez2008} identifies three additional dimensions of equity: identity, power, and access. Even if we were able to eliminate the achievement gap entirely by changing the grade scale, we still might have inequities show up in these other dimensions. Put another way, even if a course achieves Equity of Parity by eliminating achievement gaps, it might still be the case that certain student groups (on average, for example) suffer from less access to participation, or do not feel as much a part of the classroom culture as other groups. In this paper, we use the the Course Deficit Model with Equity of Parity on  student achievement as measured by grades. This is only meant to be one tool among many that we use to address larger systemic implications of racism in our classrooms.

The newest data in this study are 10 years old at the time of submission. This is useful for the analysis conducted in section \ref{sec:LaterCourses} because after this amount of time a vast majority of students in the data set have graduated or left the University so the data set is as complete as possible. However, in the past 10 years a number of things have changed. For example, many universities have launched efforts to decrease equity gaps at the course level, and recently the Covid-19 pandemic has shifted the narrative about grades in higher education and contributed to more instructors looking towards resources for equitable grading \cite{feldman2018grading, Guskey2014, Dueck2014} and in some cases experimenting with the idea of ``Ungrading'' \cite{Blum2020}.  The grades in the courses analysed in this study came almost exclusively from quiz and exam grades, and courses with different kinds of weighting would differently impact the students course grade. We caution that this limitation needs further investigation and should not be used to justify the use of the percent scale unless future data support that certain accommodations of the scale support equity of parity when using a Course Deficit Model.

As much research shows, exam scores are only a proxy for student understanding. For example, conceptual inventories are only somewhat correlated with exam scores \cite{West2006} and the ability to solve problems doesn't necessarily indicate content mastery  \cite{Kim2002Students}. Even with our best metrics for measuring student understanding there are unexplained biases (e.g. \cite{Madsen2013}). Since the fraction of A's metric is also subject to a grade penalty for URM students, more research is needed to understand the nature of this grade penalty on this metric, and what impact that might have on our analysis.   

Changing the grade scale is just one possible way of reducing inequity in a course. We still see difference in fraction of A’s and missing work in the 4.0 scale course, and this course still has a measurable equity gap. Therefore, changing the grade scale is only a partial solution to achieving Equity of Parity. The Course Deficit Model indicates that there is still work to do to make our courses equitable.

\section{ACKNOWLEDGMENTS}
We would like to thank the San Jose State University PER group for reviewing and providing feedback on an earlier draft of this paper.

\bibliography{CassBib.bib}

\begin{thebibliography}{38}%
\makeatletter
\providecommand \@ifxundefined [1]{%
 \@ifx{#1\undefined}
}%
\providecommand \@ifnum [1]{%
 \ifnum #1\expandafter \@firstoftwo
 \else \expandafter \@secondoftwo
 \fi
}%
\providecommand \@ifx [1]{%
 \ifx #1\expandafter \@firstoftwo
 \else \expandafter \@secondoftwo
 \fi
}%
\providecommand \natexlab [1]{#1}%
\providecommand \enquote  [1]{``#1''}%
\providecommand \bibnamefont  [1]{#1}%
\providecommand \bibfnamefont [1]{#1}%
\providecommand \citenamefont [1]{#1}%
\providecommand \href@noop [0]{\@secondoftwo}%
\providecommand \href [0]{\begingroup \@sanitize@url \@href}%
\providecommand \@href[1]{\@@startlink{#1}\@@href}%
\providecommand \@@href[1]{\endgroup#1\@@endlink}%
\providecommand \@sanitize@url [0]{\catcode `\\12\catcode `\$12\catcode
  `\&12\catcode `\#12\catcode `\^12\catcode `\_12\catcode `\%12\relax}%
\providecommand \@@startlink[1]{}%
\providecommand \@@endlink[0]{}%
\providecommand \url  [0]{\begingroup\@sanitize@url \@url }%
\providecommand \@url [1]{\endgroup\@href {#1}{\urlprefix }}%
\providecommand \urlprefix  [0]{URL }%
\providecommand \Eprint [0]{\href }%
\providecommand \doibase [0]{https://doi.org/}%
\providecommand \selectlanguage [0]{\@gobble}%
\providecommand \bibinfo  [0]{\@secondoftwo}%
\providecommand \bibfield  [0]{\@secondoftwo}%
\providecommand \translation [1]{[#1]}%
\providecommand \BibitemOpen [0]{}%
\providecommand \bibitemStop [0]{}%
\providecommand \bibitemNoStop [0]{.\EOS\space}%
\providecommand \EOS [0]{\spacefactor3000\relax}%
\providecommand \BibitemShut  [1]{\csname bibitem#1\endcsname}%
\let\auto@bib@innerbib\@empty
\bibitem [{\citenamefont {Webb}\ \emph {et~al.}(2020)\citenamefont {Webb},
  \citenamefont {Paul},\ and\ \citenamefont {Chessey}}]{Webb2020a}%
  \BibitemOpen
  \bibfield  {author} {\bibinfo {author} {\bibfnamefont {D.~J.}\ \bibnamefont
  {Webb}}, \bibinfo {author} {\bibfnamefont {C.~A.}\ \bibnamefont {Paul}},\
  and\ \bibinfo {author} {\bibfnamefont {M.~K.}\ \bibnamefont {Chessey}},\
  }\bibfield  {title} {\bibinfo {title} {{Relative impacts of different
  grade-scales on student success in introductory physics}},\ }\href
  {https://doi.org/10.1103/PhysRevPhysEducRes.16.020114} {\bibfield  {journal}
  {\bibinfo  {journal} {Physical Review Physics Education Research}\ }\textbf
  {\bibinfo {volume} {16}},\ \bibinfo {pages} {20114} (\bibinfo {year}
  {2020})},\ \Eprint {https://arxiv.org/abs/1903.06747} {arXiv:1903.06747}
  \BibitemShut {NoStop}%
\bibitem [{\citenamefont {Cotner}\ and\ \citenamefont
  {Ballen}(2017)}]{cotner2017}%
  \BibitemOpen
  \bibfield  {author} {\bibinfo {author} {\bibfnamefont {S.}~\bibnamefont
  {Cotner}}\ and\ \bibinfo {author} {\bibfnamefont {C.~J.}\ \bibnamefont
  {Ballen}},\ }\bibfield  {title} {\bibinfo {title} {{Can mixed assessment
  methods make biology classes more equitable?}},\ }\bibfield  {journal}
  {\bibinfo  {journal} {PLoS ONE}\ }\textbf {\bibinfo {volume} {12}},\ \href
  {https://doi.org/10.1371/journal.pone.0189610} {10.1371/journal.pone.0189610}
  (\bibinfo {year} {2017})\BibitemShut {NoStop}%
\bibitem [{\citenamefont {Valencia}(1997)}]{Valencia1997}%
  \BibitemOpen
  \bibfield  {author} {\bibinfo {author} {\bibfnamefont {R.~R.}\ \bibnamefont
  {Valencia}},\ }\href@noop {} {\emph {\bibinfo {title} {{The Evolution of
  Deficit Thinking: Educational Thought and Practice}}}}\ (\bibinfo
  {publisher} {The Falmer Press},\ \bibinfo {address} {London},\ \bibinfo
  {year} {1997})\BibitemShut {NoStop}%
\bibitem [{\citenamefont {Rodriguez}\ \emph {et~al.}(2012)\citenamefont
  {Rodriguez}, \citenamefont {Brewe}, \citenamefont {Sawtelle},\ and\
  \citenamefont {Kramer}}]{Rodriguez2012Impact}%
  \BibitemOpen
  \bibfield  {author} {\bibinfo {author} {\bibfnamefont {I.}~\bibnamefont
  {Rodriguez}}, \bibinfo {author} {\bibfnamefont {E.}~\bibnamefont {Brewe}},
  \bibinfo {author} {\bibfnamefont {V.}~\bibnamefont {Sawtelle}},\ and\
  \bibinfo {author} {\bibfnamefont {L.~H.}\ \bibnamefont {Kramer}},\ }\bibfield
   {title} {\bibinfo {title} {{Impact of equity models and statistical measures
  on interpretations of educational reform}},\ }\href
  {https://doi.org/10.1103/physrevstper.8.020103} {\bibfield  {journal}
  {\bibinfo  {journal} {Physical Review Special Topics - Physics Education
  Research}\ }\textbf {\bibinfo {volume} {8}},\ \bibinfo {pages} {020103+}
  (\bibinfo {year} {2012})}\BibitemShut {NoStop}%
\bibitem [{\citenamefont {Omi}\ and\ \citenamefont {Winant}(2015)}]{Omi2015}%
  \BibitemOpen
  \bibfield  {author} {\bibinfo {author} {\bibfnamefont {M.}~\bibnamefont
  {Omi}}\ and\ \bibinfo {author} {\bibfnamefont {H.}~\bibnamefont {Winant}},\
  }\href@noop {} {\emph {\bibinfo {title} {{Racial Formation in the United
  States}}}}\ (\bibinfo  {publisher} {Routledge},\ \bibinfo {address} {New
  York},\ \bibinfo {year} {2015})\BibitemShut {NoStop}%
\bibitem [{\citenamefont {Rodriguez}(2001)}]{Rodriguez2001}%
  \BibitemOpen
  \bibfield  {author} {\bibinfo {author} {\bibfnamefont {A.~J.}\ \bibnamefont
  {Rodriguez}},\ }\bibfield  {title} {\bibinfo {title} {{From gap gazing to
  promising cases: Moving toward equity in urban systemic reform}},\ }\href
  {https://doi.org/10.1002/tea.10005} {\bibfield  {journal} {\bibinfo
  {journal} {J. Res. Sci. Teach.}\ }\textbf {\bibinfo {volume} {38}},\ \bibinfo
  {pages} {1115} (\bibinfo {year} {2001})}\BibitemShut {NoStop}%
\bibitem [{\citenamefont {Guti{\'{e}}rrez}(2008)}]{Gutierrez2008}%
  \BibitemOpen
  \bibfield  {author} {\bibinfo {author} {\bibfnamefont {R.}~\bibnamefont
  {Guti{\'{e}}rrez}},\ }\bibfield  {title} {\bibinfo {title} {A ``gap-gazing''
  fetish in mathematics education? problematizing research on the achievement
  gap},\ }\href {https://doi.org/10.2307/40539302} {\bibfield  {journal}
  {\bibinfo  {journal} {Journal for Research in Mathematics Education}\
  }\textbf {\bibinfo {volume} {39}},\ \bibinfo {pages} {357} (\bibinfo {year}
  {2008})}\BibitemShut {NoStop}%
\bibitem [{\citenamefont {Salehi}\ \emph {et~al.}(2019)\citenamefont {Salehi},
  \citenamefont {Burkholder}, \citenamefont {Lepage}, \citenamefont {Pollock},\
  and\ \citenamefont {Wieman}}]{Salehi2019}%
  \BibitemOpen
  \bibfield  {author} {\bibinfo {author} {\bibfnamefont {S.}~\bibnamefont
  {Salehi}}, \bibinfo {author} {\bibfnamefont {E.}~\bibnamefont {Burkholder}},
  \bibinfo {author} {\bibfnamefont {G.~P.}\ \bibnamefont {Lepage}}, \bibinfo
  {author} {\bibfnamefont {S.}~\bibnamefont {Pollock}},\ and\ \bibinfo {author}
  {\bibfnamefont {C.}~\bibnamefont {Wieman}},\ }\bibfield  {title} {\bibinfo
  {title} {{Demographic gaps or preparation gaps?: The large impact of incoming
  preparation on performance of students in introductory physics}},\ }\href
  {https://doi.org/10.1103/physrevphyseducres.15.020114} {\bibfield  {journal}
  {\bibinfo  {journal} {Physical Review Physics Education Research}\ }\textbf
  {\bibinfo {volume} {15}},\ \bibinfo {pages} {20114} (\bibinfo {year}
  {2019})}\BibitemShut {NoStop}%
\bibitem [{\citenamefont {Shafer}\ \emph {et~al.}(2021)\citenamefont {Shafer},
  \citenamefont {Mahmood},\ and\ \citenamefont {Stelzer}}]{Shafer2021}%
  \BibitemOpen
  \bibfield  {author} {\bibinfo {author} {\bibfnamefont {D.}~\bibnamefont
  {Shafer}}, \bibinfo {author} {\bibfnamefont {M.~S.}\ \bibnamefont
  {Mahmood}},\ and\ \bibinfo {author} {\bibfnamefont {T.}~\bibnamefont
  {Stelzer}},\ }\bibfield  {title} {\bibinfo {title} {{Impact of broad
  categorization on statistical results: How underrepresented minority
  designation can mask the struggles of both Asian American and African
  American students}},\ }\href
  {https://doi.org/10.1103/PhysRevPhysEducRes.17.010113} {\bibfield  {journal}
  {\bibinfo  {journal} {Physical Review Physics Education Research}\ }\textbf
  {\bibinfo {volume} {17}},\ \bibinfo {pages} {010113} (\bibinfo {year}
  {2021})}\BibitemShut {NoStop}%
\bibitem [{\citenamefont {Webb}(2021)}]{Webb2021a}%
  \BibitemOpen
  \bibfield  {author} {\bibinfo {author} {\bibfnamefont {D.~J.}\ \bibnamefont
  {Webb}},\ }\href@noop {} {\bibinfo {title} {{Addendum to Concepts First
  Paper: A Student Deficit Model is Untenable in Understanding a Demographic
  Grade Gap}}},\ \bibinfo {howpublished} {e-print arxiv.org/abs/2109.09240}
  (\bibinfo {year} {2021})\BibitemShut {NoStop}%
\bibitem [{\citenamefont {Quinn}(2020)}]{Quinn2020}%
  \BibitemOpen
  \bibfield  {author} {\bibinfo {author} {\bibfnamefont {D.~M.}\ \bibnamefont
  {Quinn}},\ }\bibfield  {title} {\bibinfo {title} {{Experimental Effects of
  “Achievement Gap” News Reporting on Viewers’ Racial Stereotypes,
  Inequality Explanations, and Inequality Prioritization}},\ }\href
  {https://doi.org/10.3102/0013189X20932469} {\bibfield  {journal} {\bibinfo
  {journal} {Ed. Researcher}\ }\textbf {\bibinfo {volume} {49}},\ \bibinfo
  {pages} {482} (\bibinfo {year} {2020})}\BibitemShut {NoStop}%
\bibitem [{\citenamefont {Hazari}\ \emph {et~al.}(2007)\citenamefont {Hazari},
  \citenamefont {Tai},\ and\ \citenamefont {Sadler}}]{Hazari2007}%
  \BibitemOpen
  \bibfield  {author} {\bibinfo {author} {\bibfnamefont {Z.}~\bibnamefont
  {Hazari}}, \bibinfo {author} {\bibfnamefont {R.~H.}\ \bibnamefont {Tai}},\
  and\ \bibinfo {author} {\bibfnamefont {P.~M.}\ \bibnamefont {Sadler}},\
  }\bibfield  {title} {\bibinfo {title} {Gender differences in introductory
  university physics performance: The influence of high school physics
  preparation and affective factors},\ }\href
  {https://doi.org/https://doi.org/10.1002/sce.20223} {\bibfield  {journal}
  {\bibinfo  {journal} {Science Education}\ }\textbf {\bibinfo {volume} {91}},\
  \bibinfo {pages} {847} (\bibinfo {year} {2007})}\BibitemShut {NoStop}%
\bibitem [{\citenamefont {Kost}\ \emph {et~al.}(2009)\citenamefont {Kost},
  \citenamefont {Pollock},\ and\ \citenamefont
  {Finkelstein}}]{Kost2009Characterizing}%
  \BibitemOpen
  \bibfield  {author} {\bibinfo {author} {\bibfnamefont {L.~E.}\ \bibnamefont
  {Kost}}, \bibinfo {author} {\bibfnamefont {S.~J.}\ \bibnamefont {Pollock}},\
  and\ \bibinfo {author} {\bibfnamefont {N.~D.}\ \bibnamefont {Finkelstein}},\
  }\bibfield  {title} {\bibinfo {title} {{Characterizing the gender gap in
  introductory physics}},\ }\href
  {https://doi.org/10.1103/physrevstper.5.010101} {\bibfield  {journal}
  {\bibinfo  {journal} {Phys. Rev. ST Phys. Educ. Res.}\ }\textbf {\bibinfo
  {volume} {5}},\ \bibinfo {pages} {010101+} (\bibinfo {year}
  {2009})}\BibitemShut {NoStop}%
\bibitem [{\citenamefont {Lubienski}(2008)}]{Lubienski2008}%
  \BibitemOpen
  \bibfield  {author} {\bibinfo {author} {\bibfnamefont {S.~T.}\ \bibnamefont
  {Lubienski}},\ }\href {https://www.jstor.org/stable/40539301} {\bibinfo
  {title} {{On "gap gazing" in mathematics education: The need for gaps
  analyses}}} (\bibinfo {year} {2008})\BibitemShut {NoStop}%
\bibitem [{\citenamefont {Coates}(2014)}]{Coates2014}%
  \BibitemOpen
  \bibfield  {author} {\bibinfo {author} {\bibfnamefont {T.-N.}\ \bibnamefont
  {Coates}},\ }\bibfield  {title} {\bibinfo {title} {{Black Pathology and the
  Closing of the Progressive Mind}},\ }\href
  {https://www.theatlantic.com/politics/archive/2014/03/black-pathology-and-the-closing-of-the-progressive-mind/284523/}
  {\bibfield  {journal} {\bibinfo  {journal} {The Atlantic}\ } (\bibinfo {year}
  {2014})}\BibitemShut {NoStop}%
\bibitem [{\citenamefont {Kendi}(2019)}]{Kendi2019}%
  \BibitemOpen
  \bibfield  {author} {\bibinfo {author} {\bibfnamefont {I.~X.}\ \bibnamefont
  {Kendi}},\ }\href@noop {} {\emph {\bibinfo {title} {{How to Be an
  Antiracist}}}}\ (\bibinfo  {publisher} {One World},\ \bibinfo {address} {New
  York},\ \bibinfo {year} {2019})\BibitemShut {NoStop}%
\bibitem [{\citenamefont {Theobald}\ \emph {et~al.}(2020)\citenamefont
  {Theobald}, \citenamefont {Hill}, \citenamefont {Tran}, \citenamefont
  {Agrawal}, \citenamefont {Arroyo}, \citenamefont {Behling}, \citenamefont
  {Chambwe}, \citenamefont {Cintr\'{o}n}, \citenamefont {Cooper}, \citenamefont
  {Dunster}, \citenamefont {Grummer}, \citenamefont {Hennessey}, \citenamefont
  {Hsiao}, \citenamefont {Iranon}, \citenamefont {Jones~II}, \citenamefont
  {Jordt}, \citenamefont {Keller}, \citenamefont {Lacey}, \citenamefont
  {Littlefield}, \citenamefont {Lowe}, \citenamefont {Newman}, \citenamefont
  {Okolo}, \citenamefont {Olroyd}, \citenamefont {Peecook}, \citenamefont
  {Pickett}, \citenamefont {Slager}, \citenamefont {Caviedes-Solis},
  \citenamefont {Stanchak}, \citenamefont {Sundaravardan}, \citenamefont
  {Valdebenito}, \citenamefont {Williams}, \citenamefont {Zinsli},\ and\
  \citenamefont {Freeman}}]{Theobald2020}%
  \BibitemOpen
  \bibfield  {author} {\bibinfo {author} {\bibfnamefont {E.~J.}\ \bibnamefont
  {Theobald}}, \bibinfo {author} {\bibfnamefont {M.~J.}\ \bibnamefont {Hill}},
  \bibinfo {author} {\bibfnamefont {E.}~\bibnamefont {Tran}}, \bibinfo {author}
  {\bibfnamefont {S.}~\bibnamefont {Agrawal}}, \bibinfo {author} {\bibfnamefont
  {E.~N.}\ \bibnamefont {Arroyo}}, \bibinfo {author} {\bibfnamefont
  {S.}~\bibnamefont {Behling}}, \bibinfo {author} {\bibfnamefont
  {N.}~\bibnamefont {Chambwe}}, \bibinfo {author} {\bibfnamefont {D.~L.}\
  \bibnamefont {Cintr\'{o}n}}, \bibinfo {author} {\bibfnamefont {J.~D.}\
  \bibnamefont {Cooper}}, \bibinfo {author} {\bibfnamefont {G.}~\bibnamefont
  {Dunster}}, \bibinfo {author} {\bibfnamefont {J.~A.}\ \bibnamefont
  {Grummer}}, \bibinfo {author} {\bibfnamefont {K.}~\bibnamefont {Hennessey}},
  \bibinfo {author} {\bibfnamefont {J.}~\bibnamefont {Hsiao}}, \bibinfo
  {author} {\bibfnamefont {N.}~\bibnamefont {Iranon}}, \bibinfo {author}
  {\bibfnamefont {L.}~\bibnamefont {Jones~II}}, \bibinfo {author}
  {\bibfnamefont {H.}~\bibnamefont {Jordt}}, \bibinfo {author} {\bibfnamefont
  {M.}~\bibnamefont {Keller}}, \bibinfo {author} {\bibfnamefont {M.~E.}\
  \bibnamefont {Lacey}}, \bibinfo {author} {\bibfnamefont {C.~E.}\ \bibnamefont
  {Littlefield}}, \bibinfo {author} {\bibfnamefont {A.}~\bibnamefont {Lowe}},
  \bibinfo {author} {\bibfnamefont {S.}~\bibnamefont {Newman}}, \bibinfo
  {author} {\bibfnamefont {V.}~\bibnamefont {Okolo}}, \bibinfo {author}
  {\bibfnamefont {S.}~\bibnamefont {Olroyd}}, \bibinfo {author} {\bibfnamefont
  {B.~R.}\ \bibnamefont {Peecook}}, \bibinfo {author} {\bibfnamefont {S.~B.}\
  \bibnamefont {Pickett}}, \bibinfo {author} {\bibfnamefont {D.~L.}\
  \bibnamefont {Slager}}, \bibinfo {author} {\bibfnamefont {I.~W.}\
  \bibnamefont {Caviedes-Solis}}, \bibinfo {author} {\bibfnamefont {K.~E.}\
  \bibnamefont {Stanchak}}, \bibinfo {author} {\bibfnamefont {V.}~\bibnamefont
  {Sundaravardan}}, \bibinfo {author} {\bibfnamefont {C.}~\bibnamefont
  {Valdebenito}}, \bibinfo {author} {\bibfnamefont {C.~R.}\ \bibnamefont
  {Williams}}, \bibinfo {author} {\bibfnamefont {K.}~\bibnamefont {Zinsli}},\
  and\ \bibinfo {author} {\bibfnamefont {S.}~\bibnamefont {Freeman}},\
  }\bibfield  {title} {\bibinfo {title} {{Active learning narrows achievement
  gaps for underrepresented students in undergraduate science, technology,
  engineering, and math}},\ }\href {https://doi.org/10.1073/pnas.1916903117}
  {\bibfield  {journal} {\bibinfo  {journal} {Proceedings of the National
  Academy of Sciences of the United States}\ }\textbf {\bibinfo {volume}
  {117}},\ \bibinfo {pages} {6476–6483} (\bibinfo {year} {2020})}\BibitemShut
  {NoStop}%
\bibitem [{\citenamefont {Webb}(2017)}]{Webb2017}%
  \BibitemOpen
  \bibfield  {author} {\bibinfo {author} {\bibfnamefont {D.~J.}\ \bibnamefont
  {Webb}},\ }\bibfield  {title} {\bibinfo {title} {{Concepts first: A course
  with improved educational outcomes and parity for underrepresented minority
  groups}},\ }\href {https://doi.org/10.1119/1.4991371} {\bibfield  {journal}
  {\bibinfo  {journal} {Am. J. Phys.}\ }\textbf {\bibinfo {volume} {85}},\
  \bibinfo {pages} {628} (\bibinfo {year} {2017})}\BibitemShut {NoStop}%
\bibitem [{\citenamefont {Webb}\ and\ \citenamefont {Potter}(2021)}]{Webb2021}%
  \BibitemOpen
  \bibfield  {author} {\bibinfo {author} {\bibfnamefont {D.~J.}\ \bibnamefont
  {Webb}}\ and\ \bibinfo {author} {\bibfnamefont {W.~H.}\ \bibnamefont
  {Potter}},\ }\href@noop {} {\bibinfo {title} {{Gender-grade-gap zeroed out
  under a specific intro-physics assessment regime}}},\ \bibinfo {howpublished}
  {e-print arxiv.org/abs/2102.10451} (\bibinfo {year} {2021})\BibitemShut
  {NoStop}%
\bibitem [{\citenamefont {Simmons}\ and\ \citenamefont
  {Heckler}(2020)}]{Simmons2020}%
  \BibitemOpen
  \bibfield  {author} {\bibinfo {author} {\bibfnamefont {A.~B.}\ \bibnamefont
  {Simmons}}\ and\ \bibinfo {author} {\bibfnamefont {A.~F.}\ \bibnamefont
  {Heckler}},\ }\bibfield  {title} {\bibinfo {title} {{Grades, grade component
  weighting, and demographic disparities in introductory physics}},\ }\href
  {https://doi.org/10.1103/PhysRevPhysEducRes.16.020125} {\bibfield  {journal}
  {\bibinfo  {journal} {Physical Review Physics Education Research}\ }\textbf
  {\bibinfo {volume} {16}},\ \bibinfo {pages} {20125} (\bibinfo {year}
  {2020})}\BibitemShut {NoStop}%
\bibitem [{\citenamefont {Potter}\ \emph {et~al.}(2013)\citenamefont {Potter},
  \citenamefont {Webb}, \citenamefont {West}, \citenamefont {Paul},
  \citenamefont {Bowen}, \citenamefont {Weiss}, \citenamefont {Coleman},\ and\
  \citenamefont {{De Leone}}}]{Potter2013Sixteen}%
  \BibitemOpen
  \bibfield  {author} {\bibinfo {author} {\bibfnamefont {W.}~\bibnamefont
  {Potter}}, \bibinfo {author} {\bibfnamefont {D.}~\bibnamefont {Webb}},
  \bibinfo {author} {\bibfnamefont {E.}~\bibnamefont {West}}, \bibinfo {author}
  {\bibfnamefont {C.}~\bibnamefont {Paul}}, \bibinfo {author} {\bibfnamefont
  {M.}~\bibnamefont {Bowen}}, \bibinfo {author} {\bibfnamefont
  {B.}~\bibnamefont {Weiss}}, \bibinfo {author} {\bibfnamefont
  {L.}~\bibnamefont {Coleman}},\ and\ \bibinfo {author} {\bibfnamefont
  {C.}~\bibnamefont {{De Leone}}},\ }\href {http://arxiv.org/abs/1205.6970}
  {\bibinfo {title} {{Sixteen years of Collaborative Learning through Active
  Sense-making in Physics (CLASP) at UC Davis}}} (\bibinfo {year} {2013}),\
  \Eprint {https://arxiv.org/abs/1205.6970} {arXiv:1205.6970} \BibitemShut
  {NoStop}%
\bibitem [{\citenamefont {Paul}\ \emph {et~al.}(2017)\citenamefont {Paul},
  \citenamefont {Webb}, \citenamefont {Chessey},\ and\ \citenamefont
  {Potter}}]{Paul2017}%
  \BibitemOpen
  \bibfield  {author} {\bibinfo {author} {\bibfnamefont {C.~A.}\ \bibnamefont
  {Paul}}, \bibinfo {author} {\bibfnamefont {D.~J.}\ \bibnamefont {Webb}},
  \bibinfo {author} {\bibfnamefont {M.~K.}\ \bibnamefont {Chessey}},\ and\
  \bibinfo {author} {\bibfnamefont {W.~H.}\ \bibnamefont {Potter}},\ }\bibfield
   {title} {\bibinfo {title} {{Equity of success in CLASP courses at UC
  Davis}},\ }\href {https://doi.org/10.1119/perc.2017.pr.068} {\bibfield
  {journal} {\bibinfo  {journal} {2017 Physics Education Research Conference
  Proceedings}\ ,\ \bibinfo {pages} {292}} (\bibinfo {year}
  {2017})}\BibitemShut {NoStop}%
\bibitem [{\citenamefont {West}\ \emph {et~al.}(2013)\citenamefont {West},
  \citenamefont {Paul}, \citenamefont {Webb},\ and\ \citenamefont
  {Potter}}]{West2013Variation}%
  \BibitemOpen
  \bibfield  {author} {\bibinfo {author} {\bibfnamefont {E.~A.}\ \bibnamefont
  {West}}, \bibinfo {author} {\bibfnamefont {C.~A.}\ \bibnamefont {Paul}},
  \bibinfo {author} {\bibfnamefont {D.}~\bibnamefont {Webb}},\ and\ \bibinfo
  {author} {\bibfnamefont {W.~H.}\ \bibnamefont {Potter}},\ }\bibfield  {title}
  {\bibinfo {title} {{Variation of instructor-student interactions in an
  introductory interactive physics course}},\ }\href
  {https://doi.org/10.1103/physrevstper.9.010109} {\bibfield  {journal}
  {\bibinfo  {journal} {Phys. Rev. ST Phys. Educ. Res.}\ }\textbf {\bibinfo
  {volume} {9}},\ \bibinfo {pages} {010109+} (\bibinfo {year}
  {2013})}\BibitemShut {NoStop}%
\bibitem [{\citenamefont {Potter}\ \emph {et~al.}(2014)\citenamefont {Potter},
  \citenamefont {Webb}, \citenamefont {Paul}, \citenamefont {West},
  \citenamefont {Bowen}, \citenamefont {Weiss}, \citenamefont {Coleman},\ and\
  \citenamefont {{De Leone}}}]{Potter2014Sixteen}%
  \BibitemOpen
  \bibfield  {author} {\bibinfo {author} {\bibfnamefont {W.}~\bibnamefont
  {Potter}}, \bibinfo {author} {\bibfnamefont {D.}~\bibnamefont {Webb}},
  \bibinfo {author} {\bibfnamefont {C.}~\bibnamefont {Paul}}, \bibinfo {author}
  {\bibfnamefont {E.}~\bibnamefont {West}}, \bibinfo {author} {\bibfnamefont
  {M.}~\bibnamefont {Bowen}}, \bibinfo {author} {\bibfnamefont
  {B.}~\bibnamefont {Weiss}}, \bibinfo {author} {\bibfnamefont
  {L.}~\bibnamefont {Coleman}},\ and\ \bibinfo {author} {\bibfnamefont
  {C.}~\bibnamefont {{De Leone}}},\ }\bibfield  {title} {\bibinfo {title}
  {{Sixteen years of collaborative learning through active sense-making in
  physics (CLASP) at UC Davis}},\ }\href {https://doi.org/10.1119/1.4857435}
  {\bibfield  {journal} {\bibinfo  {journal} {American Journal of Physics}\
  }\textbf {\bibinfo {volume} {82}},\ \bibinfo {pages} {153} (\bibinfo {year}
  {2014})}\BibitemShut {NoStop}%
\bibitem [{\citenamefont {Wieman}\ and\ \citenamefont
  {Perkins}(2005)}]{Wieman2005}%
  \BibitemOpen
  \bibfield  {author} {\bibinfo {author} {\bibfnamefont {C.}~\bibnamefont
  {Wieman}}\ and\ \bibinfo {author} {\bibfnamefont {K.}~\bibnamefont
  {Perkins}},\ }\bibfield  {title} {\bibinfo {title} {{Transforming Physics
  Education}},\ }\href {https://doi.org/10.1063/1.2155756} {\bibfield
  {journal} {\bibinfo  {journal} {Physics Today}\ }\textbf {\bibinfo {volume}
  {58}},\ \bibinfo {pages} {36} (\bibinfo {year} {2005})}\BibitemShut {NoStop}%
\bibitem [{\citenamefont {{American Physical Society}}()}]{APS}%
  \BibitemOpen
  \bibfield  {author} {\bibinfo {author} {\bibnamefont {{American Physical
  Society}}},\ }\href@noop {} {\bibinfo {title} {{Underrepresented Minorities
  in Physics}}},\ \bibinfo {howpublished} {website:
  aps.org/programs/education/statistics/urm.cfm}\BibitemShut {NoStop}%
\bibitem [{\citenamefont {Paul}\ \emph {et~al.}(2018)\citenamefont {Paul},
  \citenamefont {Webb}, \citenamefont {Chessey},\ and\ \citenamefont
  {Lucas}}]{Paul2018}%
  \BibitemOpen
  \bibfield  {author} {\bibinfo {author} {\bibfnamefont {C.}~\bibnamefont
  {Paul}}, \bibinfo {author} {\bibfnamefont {D.~J.}\ \bibnamefont {Webb}},
  \bibinfo {author} {\bibfnamefont {M.~K.}\ \bibnamefont {Chessey}},\ and\
  \bibinfo {author} {\bibfnamefont {J.}~\bibnamefont {Lucas}},\ }\bibfield
  {title} {\bibinfo {title} {{Pondering zeros : Uncovering hidden inequities
  within a decade of grades}},\ }in\ \href
  {https://doi.org/10.1119/perc.2018.pr.Paul} {\emph {\bibinfo {booktitle}
  {2018 PERC Proceedings}}},\ \bibinfo {editor} {edited by\ \bibinfo {editor}
  {\bibfnamefont {A.}~\bibnamefont {Traxler}}, \bibinfo {editor} {\bibfnamefont
  {Y.}~\bibnamefont {Cao}},\ and\ \bibinfo {editor} {\bibfnamefont
  {S.}~\bibnamefont {Wolf}}}\ (\bibinfo {year} {2018})\BibitemShut {NoStop}%
\bibitem [{\citenamefont {Van~Dusen}\ and\ \citenamefont
  {Nissen}(2019)}]{VanDusen2019}%
  \BibitemOpen
  \bibfield  {author} {\bibinfo {author} {\bibfnamefont {B.}~\bibnamefont
  {Van~Dusen}}\ and\ \bibinfo {author} {\bibfnamefont {J.}~\bibnamefont
  {Nissen}},\ }\bibfield  {title} {\bibinfo {title} {{Modernizing use of
  regression models in physics education research: A review of hierarchical
  linear}},\ }\href {https://doi.org/https://doi.org/10.1002/sce.20223}
  {\bibfield  {journal} {\bibinfo  {journal} {Physical Review Physics Education
  Research}\ }\textbf {\bibinfo {volume} {15}},\ \bibinfo {pages} {020108}
  (\bibinfo {year} {2019})}\BibitemShut {NoStop}%
\bibitem [{\citenamefont {Snijders}\ and\ \citenamefont
  {Bosker}(1994)}]{Snijders1994}%
  \BibitemOpen
  \bibfield  {author} {\bibinfo {author} {\bibfnamefont {T.~A.~B.}\
  \bibnamefont {Snijders}}\ and\ \bibinfo {author} {\bibfnamefont {R.~J.}\
  \bibnamefont {Bosker}},\ }\bibfield  {title} {\bibinfo {title} {{Modeled
  Variance in Two-Level Models}},\ }\href
  {https://doi.org/10.1177/0049124194022003004} {\bibfield  {journal} {\bibinfo
   {journal} {Sociological Methods and Research}\ }\textbf {\bibinfo {volume}
  {22}},\ \bibinfo {pages} {342} (\bibinfo {year} {1994})}\BibitemShut
  {NoStop}%
\bibitem [{\citenamefont {Henderson}\ \emph {et~al.}(2004)\citenamefont
  {Henderson}, \citenamefont {Yerushalmi}, \citenamefont {Kuo}, \citenamefont
  {Heller},\ and\ \citenamefont {Heller}}]{Henderson2004Grading}%
  \BibitemOpen
  \bibfield  {author} {\bibinfo {author} {\bibfnamefont {C.}~\bibnamefont
  {Henderson}}, \bibinfo {author} {\bibfnamefont {E.}~\bibnamefont
  {Yerushalmi}}, \bibinfo {author} {\bibfnamefont {V.~H.}\ \bibnamefont {Kuo}},
  \bibinfo {author} {\bibfnamefont {P.}~\bibnamefont {Heller}},\ and\ \bibinfo
  {author} {\bibfnamefont {K.}~\bibnamefont {Heller}},\ }\bibfield  {title}
  {\bibinfo {title} {{Grading student problem solutions: The challenge of
  sending a consistent message}},\ }\href {https://doi.org/10.1119/1.1634963}
  {\bibfield  {journal} {\bibinfo  {journal} {Am. J. Phys.}\ }\textbf {\bibinfo
  {volume} {72}},\ \bibinfo {pages} {164} (\bibinfo {year} {2004})}\BibitemShut
  {NoStop}%
\bibitem [{\citenamefont {Marshman}\ \emph {et~al.}(2017)\citenamefont
  {Marshman}, \citenamefont {Sayer}, \citenamefont {Henderson},\ and\
  \citenamefont {Singh}}]{Marshman2017}%
  \BibitemOpen
  \bibfield  {author} {\bibinfo {author} {\bibfnamefont {E.}~\bibnamefont
  {Marshman}}, \bibinfo {author} {\bibfnamefont {R.}~\bibnamefont {Sayer}},
  \bibinfo {author} {\bibfnamefont {C.}~\bibnamefont {Henderson}},\ and\
  \bibinfo {author} {\bibfnamefont {C.}~\bibnamefont {Singh}},\ }\bibfield
  {title} {\bibinfo {title} {{Contrasting grading approaches in introductory
  physics and quantum mechanics: The case of graduate teaching assistants}},\
  }\href {https://doi.org/10.1103/PhysRevPhysEducRes.13.010120} {\bibfield
  {journal} {\bibinfo  {journal} {Physical Review Physics Education Research}\
  }\textbf {\bibinfo {volume} {13}},\ \bibinfo {pages} {1} (\bibinfo {year}
  {2017})}\BibitemShut {NoStop}%
\bibitem [{\citenamefont {Feldman}(2018)}]{feldman2018grading}%
  \BibitemOpen
  \bibfield  {author} {\bibinfo {author} {\bibfnamefont {J.}~\bibnamefont
  {Feldman}},\ }\href {https://books.google.com/books?id=ORhoDwAAQBAJ} {\emph
  {\bibinfo {title} {{Grading for Equity: What It Is, Why It Matters, and How
  It Can Transform Schools and Classrooms}}}}\ (\bibinfo  {publisher} {SAGE
  Publications},\ \bibinfo {year} {2018})\BibitemShut {NoStop}%
\bibitem [{\citenamefont {Guskey}(2014)}]{Guskey2014}%
  \BibitemOpen
  \bibfield  {author} {\bibinfo {author} {\bibfnamefont {T.~R.}\ \bibnamefont
  {Guskey}},\ }\href@noop {} {\emph {\bibinfo {title} {{On your mark :
  challenging the conventions of grading and reporting}}}}\ (\bibinfo
  {publisher} {Solution Tree Press},\ \bibinfo {year} {2014})\ p.\ \bibinfo
  {pages} {134}\BibitemShut {NoStop}%
\bibitem [{\citenamefont {Dueck}(2014)}]{Dueck2014}%
  \BibitemOpen
  \bibfield  {author} {\bibinfo {author} {\bibfnamefont {M.}~\bibnamefont
  {Dueck}},\ }\href@noop {} {\emph {\bibinfo {title} {{Grading smarter, not
  harder : assessment strategies that motivate kids and help them learn}}}}\
  (\bibinfo  {publisher} {ASCD},\ \bibinfo {year} {2014})\ p.\ \bibinfo {pages}
  {179}\BibitemShut {NoStop}%
\bibitem [{\citenamefont {Blum}\ \emph {et~al.}(2020)\citenamefont {Blum},
  \citenamefont {Blackwelder}, \citenamefont {Blum}, \citenamefont
  {Chiaravalli}, \citenamefont {Chu}, \citenamefont {Davidson}, \citenamefont
  {Gibbs}, \citenamefont {Katopodis}, \citenamefont {Kirr}, \citenamefont
  {Kohn}, \citenamefont {Riesbeck}, \citenamefont {Sackstein}, \citenamefont
  {Schultz-Bergin}, \citenamefont {Sorensen-Unruh}, \citenamefont {Stommel},\
  and\ \citenamefont {Warner}}]{Blum2020}%
  \BibitemOpen
  \bibfield  {author} {\bibinfo {author} {\bibfnamefont {S.~D.}\ \bibnamefont
  {Blum}}, \bibinfo {author} {\bibfnamefont {A.}~\bibnamefont {Blackwelder}},
  \bibinfo {author} {\bibfnamefont {S.~D.}\ \bibnamefont {Blum}}, \bibinfo
  {author} {\bibfnamefont {A.}~\bibnamefont {Chiaravalli}}, \bibinfo {author}
  {\bibfnamefont {G.}~\bibnamefont {Chu}}, \bibinfo {author} {\bibfnamefont
  {C.~N.}\ \bibnamefont {Davidson}}, \bibinfo {author} {\bibfnamefont
  {L.}~\bibnamefont {Gibbs}}, \bibinfo {author} {\bibfnamefont
  {C.}~\bibnamefont {Katopodis}}, \bibinfo {author} {\bibfnamefont
  {J.}~\bibnamefont {Kirr}}, \bibinfo {author} {\bibfnamefont {A.}~\bibnamefont
  {Kohn}}, \bibinfo {author} {\bibfnamefont {C.}~\bibnamefont {Riesbeck}},
  \bibinfo {author} {\bibfnamefont {S.}~\bibnamefont {Sackstein}}, \bibinfo
  {author} {\bibfnamefont {M.}~\bibnamefont {Schultz-Bergin}}, \bibinfo
  {author} {\bibfnamefont {C.}~\bibnamefont {Sorensen-Unruh}}, \bibinfo
  {author} {\bibfnamefont {J.}~\bibnamefont {Stommel}},\ and\ \bibinfo {author}
  {\bibfnamefont {J.}~\bibnamefont {Warner}},\ }\href@noop {} {\emph {\bibinfo
  {title} {{Ungrading: Why Rating Students Undermines Learning (and What to Do
  Instead) (Teaching and Learning in Higher Education)}}}},\ edited by\
  \bibinfo {editor} {\bibfnamefont {S.~D.}\ \bibnamefont {Blum}}\ (\bibinfo
  {publisher} {West Virginia University Press},\ \bibinfo {year}
  {2020})\BibitemShut {NoStop}%
\bibitem [{\citenamefont {West}(2006)}]{West2006}%
  \BibitemOpen
  \bibfield  {author} {\bibinfo {author} {\bibfnamefont {E.~A.}\ \bibnamefont
  {West}},\ }\emph {\bibinfo {title} {{Identifying the elements of physics
  courses that impact student learning : curriculum , instructor , peers , and
  assessment By}}},\ \href@noop {} {Ph.D. thesis},\ \bibinfo  {school}
  {University of California - Davis} (\bibinfo {year} {2006})\BibitemShut
  {NoStop}%
\bibitem [{\citenamefont {Kim}\ and\ \citenamefont
  {Pak}(2002)}]{Kim2002Students}%
  \BibitemOpen
  \bibfield  {author} {\bibinfo {author} {\bibfnamefont {E.}~\bibnamefont
  {Kim}}\ and\ \bibinfo {author} {\bibfnamefont {S.-J.}\ \bibnamefont {Pak}},\
  }\bibfield  {title} {\bibinfo {title} {{Students do not overcome conceptual
  difficulties after solving 1000 traditional problems}},\ }\href
  {https://doi.org/10.1119/1.1484151} {\bibfield  {journal} {\bibinfo
  {journal} {American Journal of Physics}\ }\textbf {\bibinfo {volume} {70}},\
  \bibinfo {pages} {759} (\bibinfo {year} {2002})}\BibitemShut {NoStop}%
\bibitem [{\citenamefont {Madsen}\ \emph {et~al.}(2013)\citenamefont {Madsen},
  \citenamefont {McKagan},\ and\ \citenamefont {Sayre}}]{Madsen2013}%
  \BibitemOpen
  \bibfield  {author} {\bibinfo {author} {\bibfnamefont {A.}~\bibnamefont
  {Madsen}}, \bibinfo {author} {\bibfnamefont {S.~B.}\ \bibnamefont
  {McKagan}},\ and\ \bibinfo {author} {\bibfnamefont {E.~C.}\ \bibnamefont
  {Sayre}},\ }\bibfield  {title} {\bibinfo {title} {{Gender gap on concept
  inventories in physics: What is consistent, what is inconsistent, and what
  factors influence the gap?}},\ }\href
  {https://doi.org/10.1103/PHYSREVSTPER.9.020121/FIGURES/5/MEDIUM} {\bibfield
  {journal} {\bibinfo  {journal} {Physical Review Special Topics - Physics
  Education Research}\ }\textbf {\bibinfo {volume} {9}},\ \bibinfo {pages}
  {020121} (\bibinfo {year} {2013})}\BibitemShut {NoStop}%
\end{thebibliography}%


%

\end{document}